\documentclass[3p]{elsarticle}
\journal{International Journal of Multiphase Flow}

\usepackage{lineno,hyperref}
\usepackage[title]{appendix}
\usepackage{amsmath}
\usepackage{dsfont}
\usepackage{array}
\usepackage{amssymb}
\usepackage{bm}
\usepackage{algorithm}
\usepackage{algpseudocode}
\usepackage{enumerate}
\usepackage{subcaption}
\usepackage{paralist}
\usepackage{multirow}
\usepackage{outlines}
\usepackage{todonotes}
\usepackage{tikz}
\usetikzlibrary{shapes}
\hypersetup{
	colorlinks,
	linkcolor={blue!50!black},
	citecolor={blue!50!black},
	urlcolor={blue!80!black},
	anchorcolor = {blue!80!black},
	filecolor = {blue!80!black},
	menucolor = {blue!80!black},
	runcolor = {blue!80!black}
}
\newcommand{\pdv}[2]{\frac{\partial#1}{\partial#2}}

\newcommand*{\zeroCddots}{{\ddots}\kern-0.55em\raisebox{1.4ex}{\scriptsize $0$}}
\newcommand*{\oneCddots}{{\ddots}\kern-0.55em\raisebox{1.4ex}{\scriptsize $1$}}
\newcommand*{\uCddots}{{\ddots}\kern-0.55em\raisebox{1.4ex}{\scriptsize $u$}}
\newcommand*{\UiCddots}{{\ddots}\kern-0.55em\raisebox{1.4ex}{\scriptsize $U_i$}}
\newcommand*{\vCddots}{{\ddots}\kern-0.55em\raisebox{1.4ex}{\scriptsize $v$}}
\newcommand*{\wCddots}{{\ddots}\kern-0.55em\raisebox{1.4ex}{\scriptsize $w$}}

\newcommand*{\zeroCdots}{{\dots}\kern-0.8em\raisebox{0.5ex}{\scriptsize $0$}}
\newcommand*{\zeroCvdots}{{\vdots}\kern-0.0em\raisebox{0.55ex}{\scriptsize $0$}}
\AtBeginEnvironment{bmatrix}{\setlength{\arraycolsep}{4pt}}

\newcommand{\markerone}{%
  \raisebox{0.5pt}{%
    \tikz{\node[draw=black, scale=0.4, circle](){};}%
  }%
}

\newcommand{\markerfour}{\raisebox{0.5pt}{\tikz{\node[draw,scale=0.4,regular polygon, regular polygon sides=4,fill=none](){};}}}

\DeclareRobustCommand\fullGray  {\tikz[baseline=-0.6ex]\draw[thick, white!50!black] (0,0)--(0.25,0);}
\DeclareRobustCommand\fullRed  {\tikz[baseline=-0.6ex]\draw[thick, red] (0,0)--(0.25,0);}
\DeclareRobustCommand\fullBlue  {\tikz[baseline=-0.6ex]\draw[thick, blue] (0,0)--(0.25,0);}

\DeclareRobustCommand\dashedGreen{\tikz[baseline=-0.6ex]\draw[thick,dashed,green!50!black] (0,0)--(0.35,0);}

\DeclareRobustCommand\chainBlack {\tikz[baseline=-0.6ex]\draw[thick,dash dot ,black] (0,0)--(0.5,0);}

\begin{document}

\begin{frontmatter}

%\title{LES of ECN Spray A using the $\Sigma$ model with a coupled and consistent finite-rate phase change model}
\title{An LES model with finite-rate phase change and subgrid spray based on a thermodynamically consistent four-equation multiphase model}
%\title{Extension of a thermodynamically consistent four-equation multiphase model with finite-rate phase change and subgrid spray}

\author[Stanford]{Henry Collis\corref{mycorrespondingauthor}}
\ead{hcollis@stanford.edu}
\cortext[mycorrespondingauthor]{Corresponding author}
\author[KTH,Stanford]{Shahab Mirjalili}
\ead{msey@kth.se}
\author[Stanford]{Makrand Khanwale}
\ead{khanwale@stanford.edu}
\author[Stanford]{Ali Mani}
\ead{alimani@stanford.edu}
\author[Stanford]{Gianluca Iaccarino}
\ead{jops@stanford.edu}

\address[Stanford]{Department of Mechanical Engineering, Stanford, CA 94305, USA}
\address[KTH]{FLOW, Department of Engineering Mechanics, KTH Royal Institute of Technology, SE-10044 Stockholm, Sweden}

\begin{abstract}
In this work, an LES model with finite-rate phase change and subgrid spray based on a high-resolution numerical scheme for multiphase multi-component simulations which satisfies interface equilibrium and phase immiscibility conditions \cite{collis2026robust} is proposed. The multiphase model is based on a robust implementation of the four-equation multiphase model which assumes a strict subgrid equilibrium of pressure, temperature, and velocity. Critically, the equilibrium assumptions of the four-equation model provide large computational savings compared to modeling the full non-equilibrium multiphase system. To obtain predictive capabilities with these restrictive equilibrium assumptions, a new phase-confined form of the Eulerian $\Sigma$ spray model is proposed to predict subgrid interfacial surface area while avoiding unphysical leakage across interfaces. Additionally, an improved finite rate phase change model which is thermodynamically bounded by the equilibration of the Gibbs-free energy is coupled with the $\Sigma$ equation to model complex phase change regimes. The full modeling framework is validated using the Engine Combustion Network (ECN) Spray A case in non-evaporating and evaporating conditions and shows excellent agreement with experimental measurements. 
\end{abstract}

\begin{keyword}
Subgrid spray, phase change, compressible, four-equation model, multicomponent, multiphase
\end{keyword}

\end{frontmatter}

\section{Introduction}

Modeling engineering-scale multiphase multi-component systems is critical for multiple applications, including combustion engines with fuel injection atomizers, rocket propellant injection, and fire-suppression systems. To incorporate the critical physics required by these applications, multiphysics modeling is needed that  captures intraphase mixing, high-density ratio interfaces, and mass-transfer models. In this work, a high-resolution numerical strategy for simulating multiphase multi-component flows based on the four-equation model is extended to include finite-rate phase change models and subgrid spray. Simulations involving subgrid spray physics can physically expect to have non-equilibrium between phase velocity, as the disperse and discrete phases do not need to have the same inertial properties. One of the overarching goals of this work is to showcase the extent in which the four-equation multiphase equilibrium conditions can be pushed while retaining robust and accurate multiphase simulations. Compared to the fully non-equilibrium multiphase models, the four-equation approach assumes mechanical (pressure) and thermal (temperature) equilibrium between phases. The governing system is simplified to a shared momentum and energy equation, greatly reducing the computational cost compared to the non-equilibrium models. In this work, we show that coupling the four-equation numerical framework proposed in \cite{collis2026robust} with appropriate models for subgrid spray and phase change in the four-equation setting achieves excellent agreement with experiments without requiring the full non-equilibrium multiphase system. 

\subsection{Phase Change}

Modeling the phase change process, including boiling, evaporation, flashing, and condensation, is critical in multiphase systems. In particular, engineering applications including diesel spray engines and liquid rocket propulsion systems require transfer of mass from the liquid propellant to gas to achieve successful ignition. In general, interphase mass transfer is governed by a non-equilibrium chemical potential between components in separate phases (details in Section \ref{sec:PhaseChangeModeling}). Modeling the rate at which the system reaches chemical equilibrium is a critical aspect for accurate phase change modeling. Two general assumptions exist for phase change modeling which can separate the modeling mechanism for the phase change processes listed above (e.g. evaporation vs boiling): instantaneous and finite rate equilibration of the chemical potential.

\subsubsection{Instantaneous Equilibrium} \label{sec:Intro-HEM}
Assuming an instantaneous equilibration of the chemical potential between phases is known as the Homogeneous Equilibrium Model (HEM) \cite{palacz2017hem}. The HEM model is appropriate for two regimes: fully resolved DNS where the phase change rate is governed by heat conduction (evaporation), and under-resolved spray which are governed by rapid phase change processes, including flashing and cavitation events \cite{saurel2016general,chiapolino2017simple}. For the cases between the extremes of interface resolved multiphase interfaces and substantially under-resolved simulations of flashing or cavitation, the HEM will over-predict the time-scale of mass transfer. The infinite rate relaxation of the chemical potential to equilibrium of the HEM is not applicable for interfacial structures which have a finite surface area but are not resolvable by a computational grid. These regimes could include using the HEM to model the evaporation of subgrid ligaments and droplets. 

\subsubsection{Finite-Rate Models} \label{sec:Intro-FiniteRateModels}
Finite-rate phase change models are designed to alleviate the issues of overpredicting mass transfer which occurs with the HEM. The finite-rate mass transfer time-scale is dependent on both the thermodynamic state, the composition, and the interfacial surface area. The simplest approach for modeling finite rate phase change is assuming a relaxation time-scale which is constant for all space and time throughout a simulation. This approach has been overviewed by multiple past studies \cite{pelanti2022arbitrary,demou2022pressure} which have qualitatively shown how important engineering quantities, including the vaporization mass present in a high-pressure fuel injector \cite{pelanti2022arbitrary}, are strongly impacted by the assigned phase change rate, showing that the ad-hoc approach of assigning a constant spatial-temporal phase change rate to a system is not practical for producing a predictive and generalizable computational model. The results of these works motivate the need for a predictive model of the finite rate phase change process.

The Homogeneous Relaxation Model (HRM) defines a functional dependence in space and time to determine the finite-rate evaporation rate. The original HRM model was proposed to approximate flashing experiments \cite{downar1996non}. The HRM model has been used extensively in computational studies of flashing and spray flows. These include works on flashing cryogenic nitrogen \cite{gaertner2020numerical}, water \cite{lyras2019numerical}, and include studying CO$_2$ expansion during refrigeration cycles \cite{palacz2017hem}. Although the HRM model has seen success when applied to flashing flows using fits to experimental data, general HRM models have not been proposed to capture the range of evaporation rates present in the dilute zones created by flashing sprays. Instead, droplet evaporation should follow the D2 law, which states that evaporation is proportional to the surface area (which for a droplet is proportional to the diameter squared). Similar to HRM models, correlations with experiments have been created to model droplet evaporation. A popular correlation was proposed by \cite{abramzon1989droplet} which uses film theory to model heat transfer through a droplet. 

In addition to correlation models for droplet flows, there are kinetic models \cite{hertz1882ueber,knudsen1915maximale,sazhin2006advanced} that are built on fundamental molecular assumptions and can capture a wider range of physical processes, including evaporation, condensation, and flashing. A popular model is based on the Hertz-Knudsen approach \cite{hertz1882ueber,knudsen1915maximale} and has been explored in multiple works \cite{fuster2010influence,lyras2024modelling}. Although the Hertz-Knudsen phase change model is derived in a manner to achieve the generality required to handle flows ranging from both flashing and droplet evaporation systems, a critical aspect of modeling mass transfer with the Hertz-Knudsen model is determining the interfacial surface area. 

\subsection{Spray Modeling}
Several approaches exist for modeling interfacial area at different levels of fidelity. A widely used approach is known as the Discrete Droplet Model (DDM) \cite{dukowicz1980particle} where the liquid phase is transported using Lagrangian particles that represent a grouping of droplets with identical velocity and diameter in order to reduce computational cost. In the DDM, the dynamics of the liquid are informed by the gas phase which is transported using Eulerian methods. Several studies have shown that the DDM approach is effective in capturing sprays \cite{reitz1987modeling,devassy2015atomization,xue2013large}. In general, determining the size of the droplets as well as the modeling of the dynamics, involve model tuning with experiments \cite{xue2015eulerian,desantes2016comparison,magnotti2017detailed,jia2022calibration}. 

However, modeling the liquid with an Eulerian model has been shown to converge well when enough spatial resolution is provided to represent interfaces using interface capturing methods. If fully resolved, these simulations are considered a detailed numerical simulation which is nearly representative of reality (same as direct numerical simulation, absent capturing the smallest satellite droplets) \cite{tryggvason2011direct}. Multiple works have attempted this with advanced numerical techniques, including AMR \cite{herrmann2011simulating,khanwale2022breakup}, to simulate primary atomization. In this work, larger LES grids are of interest to reduce computational costs while attempting to retain a high-fidelity prediction. In a consistent framework, the diffuse interface methodology used to capture interfaces as described in \cite{collis2026robust} can be used to represent a subgrid transport of unresolved liquid features. The surface density approach ($\Sigma$) involves adding a transport equation to the governing equation which represents the surface area of the liquid \cite{vallet1999modelisation,burluka2001development,lebas2005coupling,lebas2009numerical,anez2019eulerian}. In an LES setting, this transport equation represents the subgrid liquid content which is not resolvable by the mesh. Multiple studies have shown that Eulerian $\Sigma$ models can achieve reasonable agreement with experiments in dense spray regions \cite{desantes2016comparison,desantes2016coupled,pandal2017computational,desantes2020eulerian}. Furthermore, the models for dense spray can either be combined with Lagrangian spray models (ELSA) to capture the dilute zone in sprays, or additional Eulerian models can be designed to predict the dilute spray. 

The $\Sigma$ models have been successfully applied in both RANS and LES contexts \cite{lebas2009numerical,pandal2017computational,lyras2019numerical,anez2019eulerian}. In the RANS context \cite{lebas2009numerical}, all of the breakup regimes are modeled with the spray model. In the LES context, hybrid approaches have been explored that use interface capturing schemes in resolvable regions of the spray, and the $\Sigma$ spray model for the unresolved regions \cite{anez2019eulerian,nykteri2021droplet}. Furthermore, LES studies have been completed in which the primary spray features are fine enough to justify using a $\Sigma$ spray model for all scales \cite{desantes2020eulerian,gaballa2023modeling}. 

\subsection{Outline}
This work extends the system described in \cite{collis2026robust} to include additional multi-physics effects. This starts by generalizing the system of equations from an interface resolved multiphase system to include sub-grid scale models for mass, momentum, and liquid spray. Additionally, phase change capabilities are added to model the transfer of mass required to reach thermo-chemical equilibrium. The additional modeling capabilities are derived in a manner consistent with the phase constrained species diffusion model described in \cite{collis2026robust}. Section \ref{sec:LES} overviews the filtered governing equations used for LES simulation. Section \ref{sec:NumericalMethod} summarizes the baseline numerical method where more details can be found in \cite{collis2026robust}. Section \ref{sec:PhaseChangeModeling} describes the basis for phase change modeling and the proposed implementation of a thermodynamically bounded finite-rate phase change model. Section \ref{sec:SprayModeling} covers the sub-grid model for interfacial surface area used to inform the finite-rate phase change approach. Section \ref{sec:SprayAResults} presents results for a non-evaporating and an evaporating Spray A case, and show that both match well with the experimentally reported results. Finally, section \ref{sec:Conclusion} provides conclusions and future outlook.

\section{Large Eddy Simulation} \label{sec:LES}

In order to extend the capabilities of the solver to handle engineering scale problems while retaining high-fidelity accuracy, large eddy simulations (LES) will be used in this work. To obtain the system of equations used in LES the compressible multiphase multi-component governing equations can be filtered and Favre averaged. In this work, the filter is represented by the grid, and the Favre averaging operator applied to a general field $f$ is given by,
\begin{equation}
    \tilde{f}=\frac{\overline{f\rho}}{\bar{\rho}}.
\end{equation}
After applying a filter and Favre averaging to the governing equations (and ignoring the subgrid content from the viscous and interface regularization terms) the system can be written as,
\begin{align}
    \pdv{\bar\rho \widetilde Y_p^c}{t} + \pdv{\widetilde u_j \bar\rho \widetilde Y_p^c}{x_j} &=
    -\pdv{}{x_j} \Big(\overline{J_{p,j}^c} + \overline{J_{p,j}^c}^t + \overline{R_{p,j}^c} \Big) + \overline{\dot m_p^c}  + \overline{\dot{\omega}^c_p}\label{eq:LESConsMass}\\
    \pdv{\bar\rho \widetilde u_i}{t} + \pdv{\widetilde u_j \bar\rho \widetilde u_i}{x_j} &=
    -\pdv{}{x_j} \Big(\overline{P}\delta_{ij} + \overline{\tau_{ij}} + \overline{\tau_{ij}}^t + \widetilde u_i \overline{R_j} \Big) \label{eq:LESConsMom}\\
    \pdv{\bar\rho \widetilde E}{t} + \pdv{\widetilde u_j \bar\rho \widetilde E}{x_j} &=
    -\pdv{}{x_j} \Big(\overline{P} \widetilde {u_j} + \overline{\tau_{ij}}\widetilde{u_j} + \overline{\tau_{ij}}^t\widetilde{u_j} + \overline{Q_j} + \overline{Q_j}^t + \overline{H_j} + \widetilde{u_i} \widetilde{u_i} \overline{R_j}/2 \Big) \label{eq:LESConsEng}
\end{align}
where, to clearly distinguish between different phases, the phase of a material will be indicated by $p$, where the subscript $p = 1,2$, and the components within a given phase are indicated by the superscript $c$, where $1 \leq c \leq N$. Index notation will be used for coordinates and tensor components, although we do not imply index notation for $p$ and $c$. The definition for the primitive variables includes $Y^c_p$ as the mass of component $c$ of phase $p$ per total mass, $u_i = [u,v,w]^T$ as the velocity vector, $P$ as the pressure, $T$ as the temperature, $\rho$ as the mixture density, and $\rho E$ as the total energy per unit volume defined as $\rho E = \rho e + \frac{1}{2}\rho u_i u_i$ where $e$ is the internal energy per unit volume.
 
To model the sub-grid content present in an LES setting, sub-grid stresses have been added to the right-hand-side of the system and are indicated with superscript $t$. For instance, the combined viscous stress tensor and the subgrid momentum flux, $\overline{\tau_{ij}}+\overline{\tau_{ij}}^t$, are defined as,
\begin{equation}
    \overline {\tau_{ij}} + \overline {\tau_{ij}}^t = -(\overline{\mu} + \mu^t)\left[\frac{\partial \widetilde u_i}{\partial x_j} + \frac{\partial \widetilde  u_j}{\partial x_i} - 2/3\frac{ \partial \widetilde u_k}{\partial x_k}\delta_{ij}\right]
\end{equation} 
and $\overline{Q_i}+\overline{Q_i}^t$ is defined as,
\begin{equation}
    \overline Q_i + \overline Q_i^t = -(\overline \lambda+\lambda^t) \frac{\partial \widetilde T}{\partial x_i} + \sum_p\sum_c (\overline{J_{p,i}^c}+\overline{J_{p,i}^c}^t) \frac{\widetilde Y_p^c}{\widetilde Y_p}\widetilde h_p^c
\end{equation}
where $\lambda$ is the heat conductivity of the mixture and $h_p^c$ is the specific enthalpy of component $c$ of phase $p$.
Following the model used for the species mass diffusion in \cite{collis2026robust}, the filtered species diffusion along with the subgrid mass diffusion term is given by,
\begin{equation}
    \label{eq:LESSpeciesMassDiffusion}
    \overline{J_{p,i}^c} + \overline{J_{p,i}^c}^t = - \overline{\rho}  \Bigg[\left(\overline{D^c_p} + {(D^c_p)}^t\right)\widetilde Y_p\frac{W_p^c}{W_p} \frac{\partial }{\partial x_i}  \left(\frac{\widetilde X_p^c}{\widetilde X_p}\right) - \widetilde Y_p^c\sum_j \left(\frac{W_p^j}{W_p}\right) \left(\overline{D^j_p} + (D^j_p)^t\right)\frac{\partial }{\partial  x_i} \left(\frac{\widetilde X_p^j}{\widetilde X_p}\right)\Bigg] 
\end{equation}
where $X_p^c$ is the mixture molar fraction of component $c$ of phase $p$, $X_p$ is the mixture molar fraction of phase $p$, $Y_p$ is the mixture mass fraction of phase $p$, $W_p^c$ is the molecular weight of component $c$, $W_p$ is the molecular weight of phase $p$, and $D^c_p$ is the mass diffusivity of component $c$ within phase $p$.

The subgrid terms are closed using a constant $\sigma$-model \cite{nicoud2011using} for $\mu^t$, $\lambda^t$ is modeled assuming a constant turbulent Prandlt number, $\overline{C_p}\mu^t/\lambda^t=0.7$, and $D_p^c$ is modeled assuming a constant turbulent Schmidt number, $\mu^t/(\overline{\rho}D_p^c)=0.7$. The mixing rules and calculations for all thermodynamic quantities and equations of state follow what was described in \cite{collis2026robust}. To model the subgrid mixing content of sprays, the turbulent liquid flux is closed using,
\begin{equation}
    \overline{R_{p,j}^c}=-\frac{\widetilde Y_p^c}{\widetilde Y_p}\overline \rho_p\frac{\nu_t}{Sc_t}\frac{\partial \widetilde \phi_p}{\partial x_j}
\end{equation}
where for resolved interfaces, the liquid flux can be closed using an interface regularization model, as explored in past work \cite{collis2026robust}. In addition to modeling the subgrid content for species transport, momentum, total energy, and spray, a term in Eq. \ref{eq:LESConsMass} to model the mass transfer from phase change, $\overline{\dot{m_p^c}}$, is added to the system. The phase change modeling is discussed in Section \ref{sec:PhaseChangeModeling}.

\section{Numerical Method} \label{sec:NumericalMethod} 

The spatial discretization is a hybridization of a high-order kinetic energy and entropy preserving skew-symmetric scheme \cite{kennedy2008reduced,Kuya2021,Jain2022KEEP} with a high-resolution Godunov scheme for sharp gradients described in \cite{collis2026robust}. The Godunov scheme used for gradient regions is based on essentially non-oscillatory interpolations designed to satisfy the interface equilibrium conditions across material interfaces and reconstructs numerical fluxes using an approximate HLLC Riemann solver \cite{collis2026robust}. To avoid unphysical numerical states, a positivity-preserving check is used to locally switch to a first-order reconstruction to guarantee a robust scheme \cite{wong2022positivity}. The baseline high-order hybrid numerical scheme provides low-dissipation in smooth regions and high-resolution capturing of sharp gradients without introducing spurious numerical oscillations \cite{collis2025robust}. The implementation details of the hybrid scheme are provided in \ref{sec:HybridApproach}. The system of equations is implemented in generalized curvilinear coordinates for general application on both Cartesian, rectilinear, and generalized curvilinear grids \cite{Collis2026DiffuseInterfaceCurvilinear}. The time-integration is an explicit third-order strong stability preserving Runge Kutta scheme \cite{gottlieb2001strong}. For all simulations in this work, a temporal CFL based on advection and diffusion timescales of 0.5 is used. Finally, Navier-Stokes characteristic boundary conditions (NSCBC) are used for enforcing non-reflective inflow and outflow boundary conditions \cite{thompson1990time,poinsot1992boundary,okong2002consistent,peden2023numerical,collis2025robust}.  

\section{Phase Change Modeling} \label{sec:PhaseChangeModeling} 

The following section overviews the modeling assumptions to add phase change to an existing hydrodynamic solver. When a multiphase multi-component mixture is at thermo-chemical equilibrium no mass transfer can occur. In terms of thermodynamic quantities, thermo-chemical equilibrium for an isolated system is reached when $dS = 0$ where $S = S(U,V, N_{ij})$ is the mole (N) weighted entropy for the mixture given by,

\begin{equation} \label{eq:MolalEntropy}
    dS = \left(\frac{\partial S}{\partial U}\right)_{V,N_{ij}}dU + \left(\frac{\partial S}{\partial V}\right)_{U,N_{ij}}dV + \sum^c_{i=1}\sum^p_{j=1}\left(\frac{\partial S}{\partial N_{ij}}\right)_{U,V,N_{i\neq j}}dN_{ij}
\end{equation}
where $U$ is the molal internal energy and $V$ is the molal internal volume. As $U$, $V$, and $N_{ij}$ are independent, each derivative term must individually be zero under equilibrium. For clarity of explanation, consider a pure two-phase substance ($c=1$, $p=2$) in a closed system. For the first term, the change in entropy during phase exchange can be described as,
\begin{equation}
    dS = \left(\frac{\partial S_1}{\partial U_1}\right)_{V,N_{ij}}dU_1 + \left(\frac{\partial S_2}{\partial U_2}\right)_{V,N_{ij}}dU_2.
\end{equation}
At equilibrium, $dS = 0$, and after we note that energy stays constant in the closed system -- implying that $U_1 + U_2 = $ constant -- we can note that $dU_1 = -dU_2$ and 
\begin{equation}
    \left(\frac{\partial S_1}{\partial U_1}\right)_{V,N_{ij}}dU_1 = \left(\frac{\partial S_2}{\partial U_2}\right)_{V,N_{ij}}dU_1.
\end{equation}
Finally, with $\left(\frac{\partial S}{\partial U}\right)_{V,N_{ij}} = 1/T$, we can observe that reaching equilibrium $(dS = 0)$ requires $T_1 = T_2$. Or, in other words, the system reaches thermal equilibrium. Similarly for the second term, since $V_1 + V_2$ is constant, and $\left(\frac{\partial S}{\partial V}\right)_{U,N_{ij}} = (P/T)$, using the identical analysis as above we can obtain the additional requirement that at equilibrium $P_1 = P_2$, i.e. mechanical equilibrium. The first two constraints are assumed in the formulation of the four-equation multiphase model which is used in this work. In other models, like the seven equation multiphase model \cite{baer1986two}, the thermal and mechanical equilibrium between phases would need to be reached as part of the phase change process. Instead, in this work, only the relaxation of the final term in Eq. \ref{eq:MolalEntropy} must be captured to model phase change.

The third term in Eq. \ref{eq:MolalEntropy} can be analyzed by recognizing,
\begin{equation}
    \left(\frac{\partial S}{\partial N_{i}}\right)_{U,V,N_{j}} = \frac{\mu_i}{T}
\end{equation}
where the chemical potential $\mu_i$ is given by,
\begin{equation} \label{eq:ChemicalPotential}
    \mu_i \equiv \left(\frac{\partial U}{\partial N_i}\right)_{S,V,N_j} = \left(\frac{\partial G}{\partial N_i}\right)_{T,P,N_j} = \left(\frac{\partial H}{\partial N_i}\right)_{S,P,N_j} = \left(\frac{\partial A}{\partial N_i}\right)_{T,V,N_j}
\end{equation}
with $G$ as the Gibbs free-energy, $H$ as the enthalpy, and $A$ as the Helmoltz free-energy. Therefore, at thermo-chemical equilibrium for a one component two phase mixture we can write,
\begin{equation}
    \mu_1dN_1 + \mu_2dN_2 = 0.
\end{equation}
From conservation of mass we can say $d(N_1 + N_2) = 0$ so $dN_1 = -dN_2$. So, in equilibrium, $\mu_1 = \mu_2$. When the chemical potential is not in equilibrium,
\begin{equation}
    d(S_1 + S_2) \geq 0 \Rightarrow \left(\frac{-\mu_1}{T}\right)dN_1 + \left(\frac{-\mu_2}{T}\right)dN_2 \geq 0
\end{equation}
where if we combine with the conservation of mass equation we can show,
\begin{equation}
    dN_1 + dN_2 = 0 \Rightarrow \frac{1}{T}(\mu_1 - \mu_2)dN_2 \geq 0.
\end{equation}
So, if $\mu_1 > \mu_2 \Rightarrow dN_2 > 0$, and material moves from phase 1 to phase 2. Conversely, if $\mu_2 > \mu_1 \Rightarrow dN_1 > 0$, and material moves from phase 2 to phase 1. From this analysis, we can see that reaching equilibrium with the chemical potential is the driving force of mass exchange between phases. 

With the concept of chemical potential driving phase change, we can see from the definition of chemical potential in Eq. \ref{eq:ChemicalPotential} for a specified temperature and pressure for all phases (which is naturally enforced with the four-equation model) the chemical potential is defined in terms of the Gibbs free energy, $G$, where $G = \sum_i g_iN_i$ and $g_i = h_i - s_iT$. Therefore, a general form for $\overline{\dot{m_p^c}}$ in the four-equation context is,
\begin{equation}
    \dot{m}^c_p = \nu (g^c_{m\neq p} - g^c_p)
\end{equation}
where $\nu = \nu(P, T, Y^c_p, \Sigma)$ contains a chemical relaxation inverse timescale and is dependent on the thermodynamic state as well as the interfacial area between phases, $\Sigma$. For more general applicability and ease of implementation in a general CFD solver, an equivalent model in the LES setting can be defined as,
\begin{align}
    &\overline{\dot{m_p^c}} = \frac{1}{\tau}\rho \left(\widetilde Y{_{eq}}_{p}^c - \widetilde Y_p^c\right)
\end{align}
where $Y{_{eq}}_{p}^c$ is the mass fraction of component $c$ in phase $p$ when the system has reached thermo-chemical equilibrium. The following sections overview two approaches to modeling the timescale $\tau$. Section \ref{sec:HEM} provides a computational approach if we assume that $\tau\Rightarrow 0$, and Section \ref{sec:FiniteRatePhaseChange} provides a methodology for determining a finite value for $\tau$.

\subsection{Homogeneous Equilibrium Model} \label{sec:HEM}
As described in Section \ref{sec:Intro-HEM}, a popular approach for modeling phase change is assuming an infinite rate relaxation term for the Gibbs free-energy time-scale ($\nu \Rightarrow \infty$ or $\tau \Rightarrow 0$). Using this assumption is known as the Homogeneous Equilibrium Model (HEM). The infinite relaxation rate assumption of the HEM allows for multiple paths for numerically reaching a thermo-chemical equilibrium state. For example, numerical methodologies, including exact iterative procedures \cite{saurel2016general}, or approximate solvers \cite{chiapolino2017simple}, have been studied. In this work, a fast approximate algorithm is used to determine the equilibrium state. The strategy avoids an expensive Newton-Raphson iterative solver and instead uses an algebraic update which approximates the thermo-chemical state and converges over multiple simulation time-steps.

\subsubsection{Approximate HEM Solver}
The approximate HEM approach relaxes the system using a UV-flash approach. In a UV-flash algorithm, both the mixture internal energy and the mixture specific volume stay constant over the phase change process. During the phase change process, no energy or mass leaves the computational cell and the equilibrium mixture pressure and temperature change to account for the mass transfer. The approximate algorithm proposed by \cite{chiapolino2017simple} to find an estimate for the mass fraction of component $c$ in the gas phase $g$ after phase change is described below,
\begin{enumerate}
    \item Estimate updated mass fraction to satisfy mass conservation for all $(P,T)$ using,
        \begin{equation}
            Y^c_{g}{_{_m}} = \frac{1/\rho - 1/\rho_g(P,T)}{1/\rho_g^c(P,T) - 1/\rho_l^c(P,T)}
        \end{equation}
        where,
        \begin{equation}
            1/\rho_g = \left(1 - \sum_{k\neq c}^NY_g^k\right)\left(1/\rho_l^c(P,T)\right) + \sum_{k\neq c}^NY_g^k1/\rho_g^k(P,T)            
        \end{equation}
        and the system above is evaluated at the conditions,
        \begin{equation}
            P = P, \text{ and } T = T_{sat}(P X_g^c/X_g)        
        \end{equation}
    \item Estimate updated mass fraction to satisfy energy conservation for all $(P,T)$ using,
        \begin{equation}
            Y^c_{g}{_{_e}} = \frac{e - e_g(P,T)}{e_g^c(P,T) - e_l^c(P,T)}
        \end{equation}
        where,
        \begin{equation}
            e_g = \left(1 - \sum_{k\neq c}^NY_g^k\right)e_l^c(P,T) + \sum_{k\neq c}^NY_g^ke_g^k(P,T)            
        \end{equation}
        and the system above is evaluated at the conditions,
        \begin{equation}
            P = P, \text{ and } T = T_{sat}(P X_g^c/X_g)        
        \end{equation}
    \item Estimate the updated mass fraction to satisfy chemical equilibrium using,
        \begin{equation}
             Y_g^{c}{_{sat}} = \frac{P_{sat}W_c}{P-P_{sat}(T)}\sum_{k\neq c}Y_g^k/W_k
        \end{equation}
        where the system above is evaluated at the original pressure and temperature condition and the definition for $P_{sat}$ is defined in Eq. \ref{eq:AntoineEq}.
    \item Check bound states
    \begin{enumerate}
        \item if $(Y_g^{c}{_{_m}} - Y_g^c)(Y_g^{c}{_{_e}} - Y_g^c) < 0$ or if $(Y_g^{c}{_{_m}} - Y_g^c)(Y_g^{c}{_{sat}} - Y_g^c) < 0$ no phase change occurs. Otherwise, move on to the next step.
    \end{enumerate}
    \item Take the estimate for the new $Y_p^c$ which has the smallest variation for the original using,
        \begin{equation}
            Y_g^c = Y_g^c + \text{sign}(Y_g^{c}{_{_m}} - Y_g^c)\left(\min \{|Y_g^{c}{_{_m}} - Y_g^c|,|Y_g^{c}{_{_e}} - Y_g^c|,|Y_g^{c}{_{sat}} - Y_g^c|\}\right)
        \end{equation}
\end{enumerate}
Once the system is at the equilibrium state, all mass fraction estimates from the algorithm above are identical. The saturation state is determined based on a fitted Antoine equation given by,
\begin{equation} \label{eq:AntoineEq}
    P_{sat}(T) = 10^{A - \frac{B}{C+T}}
\end{equation}
where the parameters, $A$, $B$, and $C$ are found in the NIST database \cite{linstrom1997nist}. 

To evaluate the chemical equilibrium state in the current work, a time-splitting approach is used. During time-integration, at the end of each SSP-RK3 sub-step (after the conserved variables have been updated in time including advection and diffusion processes), the mass transfer from phase change is added to the conserved variables. Since the phase change algorithm is a UV-flash algorithm, the total momentum and energy in the system are unchanged, and only the species mass equations involved in the phase change algorithm must be updated.

\subsubsection{HEM Verification}
The implementation of the HEM solver can be tested on multiple 1-D shock-tube simulations that are designed to be sensitive to the thermodynamic composition. These tests do not contain resolved phase interfaces, and instead represent a homogeneous mixture going through a phase change process, which is indicative of an under-resolved multiphase flow scenario common in spray applications. Additionally, in these tests the liquid-gas interface is not regularized using the CDI model as it has been in \cite{collis2026robust}. Recall, that for the four-equation multiphase model the volume fraction is implicitly determined as a function of the local thermodynamic state. When phase change is active, the mass transfer will determine the interface location by relaxing towards thermo-chemical equilibrium. Regularizing the phase interface using a phase field method can move mass away from the thermo-chemical equilibrium location and we hypothesis that these differing dynamics between CDI and the HEM can create an incompatibility. Attempting to solve for the interface shape with both the phase chance and the CDI model can create an unphysical competition between interface regularization towards a $\tanh$ profile, and relaxing interface towards achieving thermo-chemical equilibrium. The competition could potentially be avoided by adding a condition to the regularization terms described in \cite{collis2026robust} to only activate in multiphase zones near thermo-chemical equilibrium, though this theory has not been tested. Future work is required to develop a phase field method that is intrinsically compatible with both the four-equation model and relaxation toward thermo-chemical equilibrium during a phase change process. 

The following section will verify the implementation of the HEM phase change algorithm using a code-to-code verification test with a reference from \cite{deng2020diffuse}. As this is a code-to-code test, the same material parameters used in \cite{deng2020diffuse} are used here. For the phase change shock-tube simulations in this work, a spatial resolution of (400 x 1) was used with an advection CFL = 0.5. 

\begin{figure}[hbt!]
\begin{center}
\includegraphics[width=0.7\textwidth]{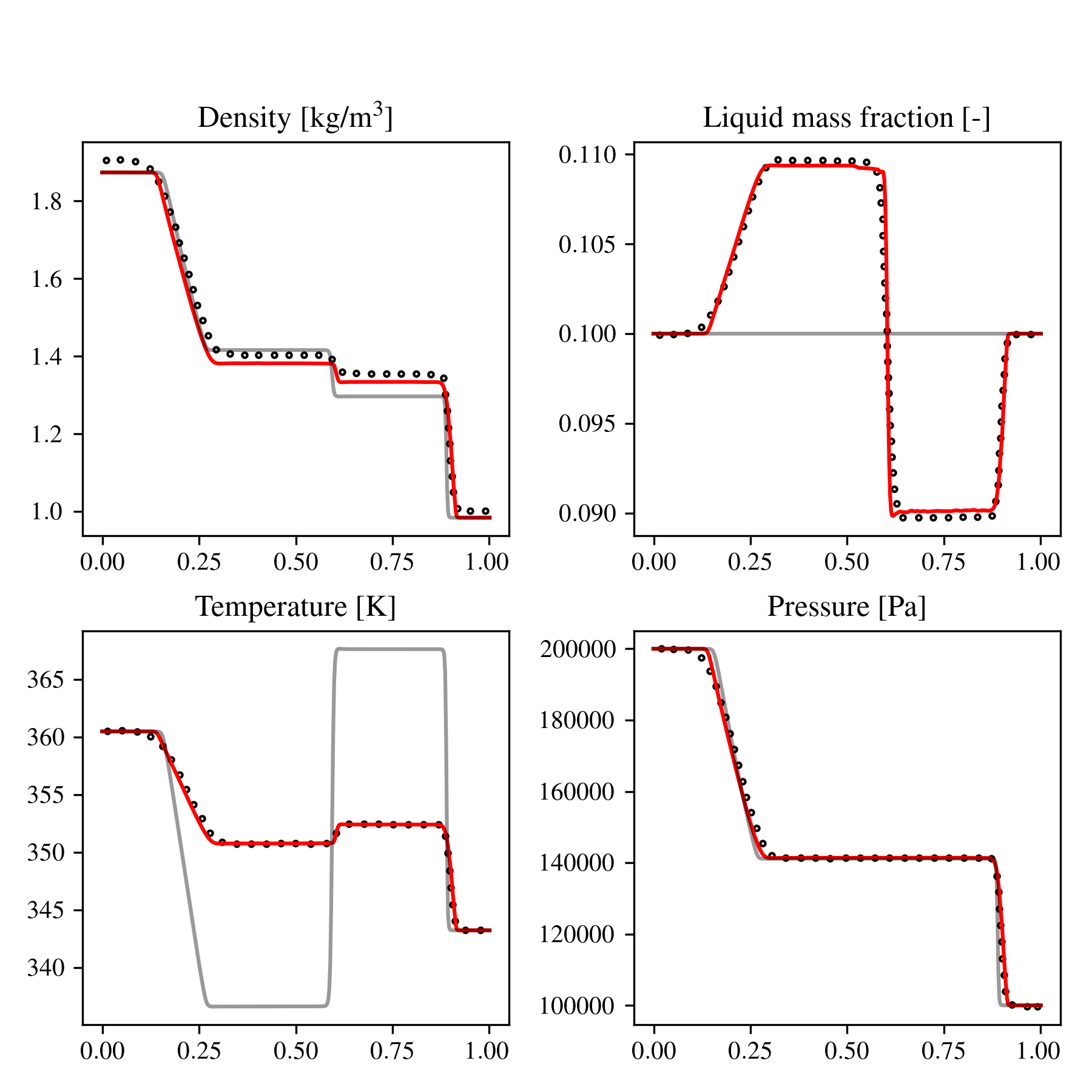}
\caption{Water-air phase change shock tube. With HEM phase change (\protect\fullRed{}), without phase change (\protect\fullGray{}) and reference with phase change \cite{deng2020diffuse} (\protect\markerone{}) \label{fig:WaterAirPhaseChange}}
\end{center}
\end{figure}

The first test starts with an air-water mixture which is far from a phase change boundary. The initial mass fractions are liquid water with $Y_l^{water} = 0.1$, $Y_g^{water}=0.2$, and $Y_g^{air}=0.7$ uniformly distributed throughout the shock-tube. Across the shock located at $x=0.5$ in a domain that ranges from $0\leq x \leq 1$, a pressure jump of $P_L = 0.2$ MPa to $P_R=0.1$ MPa is present. The temperature and density are set in order to have the mixture in thermo-chemical equilibrium, and the velocity is initially zero in the domain. For this mixture, thermo-chemical equilibrium results in $\rho_L \simeq 1.874$ kg/m$^3$, $\rho_R \simeq 0.984$ kg/m$^3$, $T_L \simeq 360.48$K, and $T_R \simeq 343.22$K. The results compared with the published work from \cite{deng2020diffuse} are shown in Figure \ref{fig:WaterAirPhaseChange} at $t=1$ms. As shown, the results agree well between the two solvers. As this mixture is far from the phase change boundary, the results without phase change (given by the gray line: \protect\fullGray{}) show a large difference compared to the phase change location in multiple fields, including the temperature and liquid mass fraction. In this case, the generality of the HEM phase change approach is showcased as both evaporation and condensation are correctly predicted.

\begin{figure}[hbt!] 
\begin{center}
\includegraphics[width=0.7\textwidth]{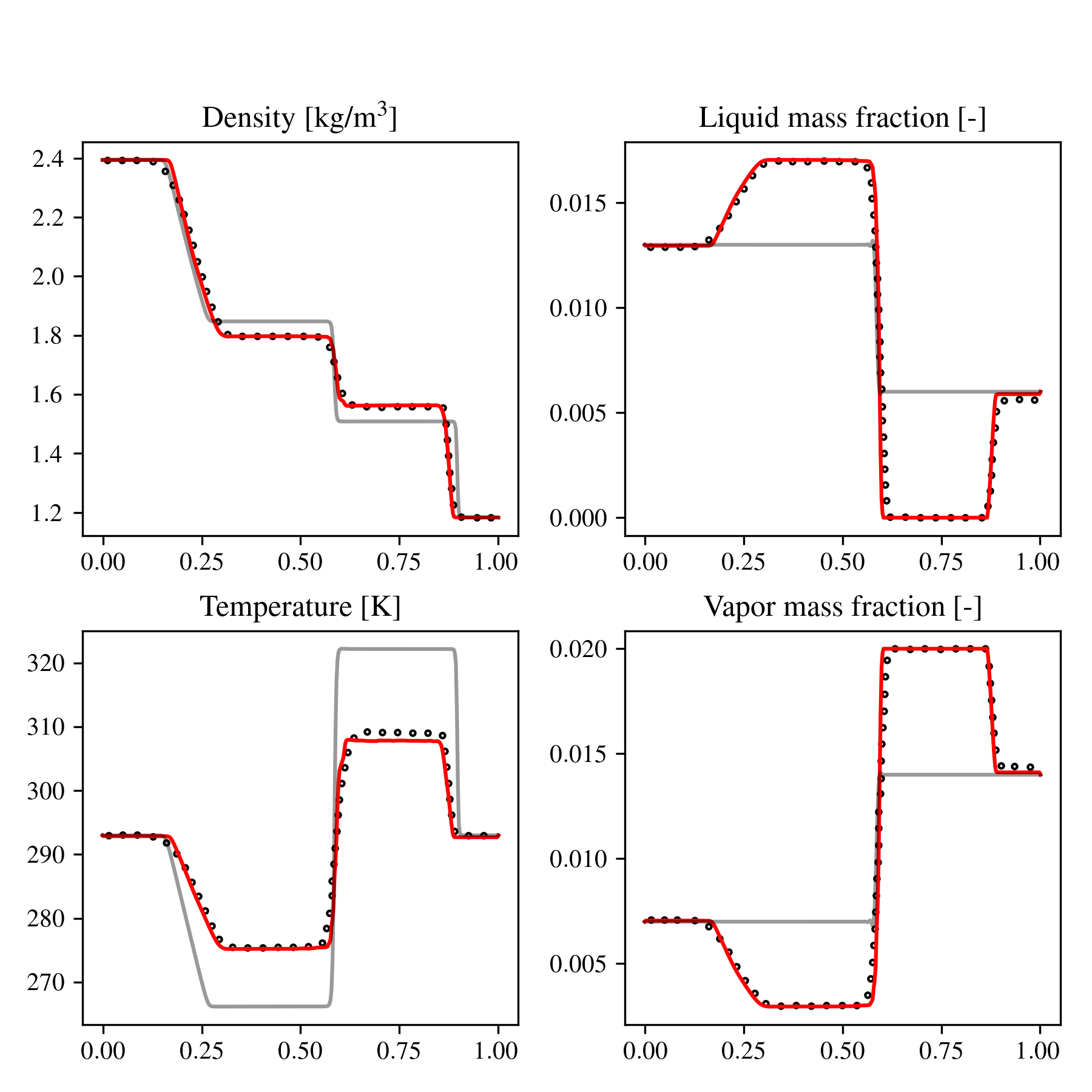}
\caption{Air-water phase change shock tube. With HEM phase change (\protect\fullRed{}), without phase change (\protect\fullGray{}), and reference with phase change \cite{deng2020diffuse} (\protect\markerone{}).\label{fig:AirWaterPhaseChange}}
\end{center}
\end{figure}

The second case used to verify the implementation of the HEM phase change solver is a shock in an air dominated mixture. The initial temperature is set to $T=293$K throughout the domain and the same initial pressure ratio of 2 from the previous case is used. In this case, the mass fraction $Y_g^{air}=0.98$ everywhere in the domain, and the mass fractions of liquid water and water vapor are deduced from satisfying thermo-chemical equilibrium with the initial condition. This results in $Y{_l^{water}}_L = 0.013$, and $Y{_l^{water}}_R = 0.006$. Figure   
\ref{fig:AirWaterPhaseChange} compares the results from the current work with \cite{deng2020diffuse} at $t=1$ ms. Similar to the first case in Figure \ref{fig:WaterAirPhaseChange}, there is excellent agreement between the codes, even though different numerical schemes are used for the spatial discretization. The effect of including phase change is still apparent, as the temperature field has a strong dependence on the application of the HEM solver. Furthermore, the liquid water experiences both evaporation and condensation, showcasing the capabilities of the consistent application of thermo-chemical relaxation to predict complex phase change.

\begin{figure}[hbt!] 
\begin{center}
\includegraphics[width=0.7\textwidth]{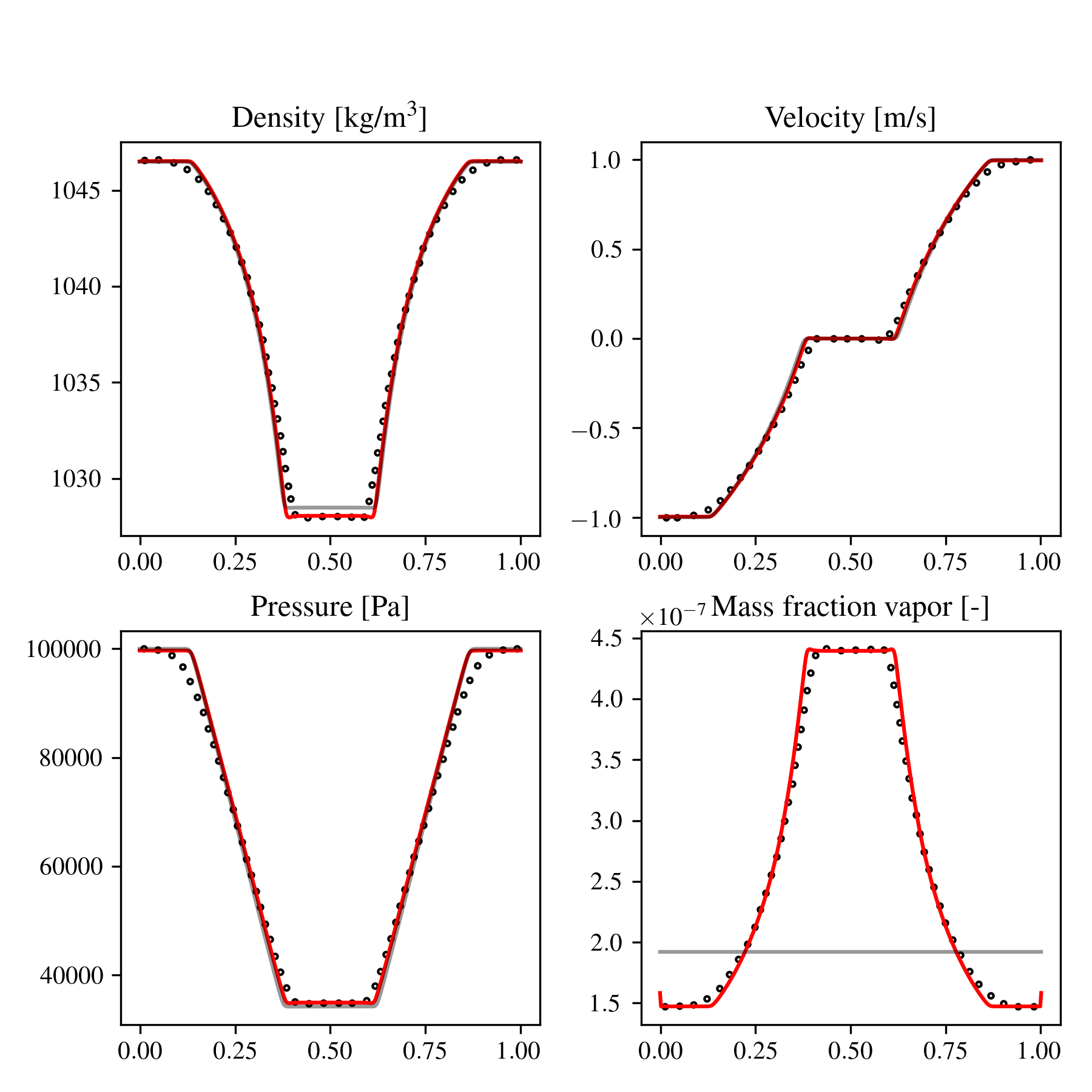}
\caption{Water expansion (cavitation) phase change shock tube. With HEM phase change (\protect\fullRed{}), without phase change (\protect\fullGray{}), and reference with phase change \cite{deng2020diffuse} (\protect\markerone{}).\label{fig:WaterExpansion}}
\end{center}
\end{figure}

The final case used to verify the ability of the HEM phase change solver to find the thermo-chemical equilibrium state is a 1D cavitation problem. The initial state involves a constant pressure, temperature, and density field at thermo-chemical equilibrium. The pressure is at 1MPa, temperature is at 293K, and density is defined using the equation of state. The velocity field is set with a positive velocity in the right of the domain of 1 m/s, and a negative velocity in the left (expansion fan) of -1 m/s. During the simulation, the velocity field induces a low pressure zone in the center of the domain and a cavitation event produces vapor from the pure liquid. The thermodynamic parameters for determining the phase change saturation state are obtained from \cite{deng2020diffuse}, but in this case the mixture is defined using the NASG stiffened gas parameters from \cite{chiapolino2017simple} to obtain results that are consistent with past works \cite{chiapolino2017simple,deng2020diffuse}. As shown in Figure \ref{fig:WaterExpansion}, the predicted vapor mass fraction results match very well with the reference solution \cite{deng2020diffuse} at t=3.5 ms, providing confidence in the current implementation of the HEM algorithm.  

\subsection{Finite-Rate Phase Change} \label{sec:FiniteRatePhaseChange}

As described in Section \ref{sec:HEM}, the assumptions of an infinite rate relaxation of the chemical potential to equilibrium are not physical for interfacial structures which have a finite surface area but are not resolvable by a LES grid. The current section will briefly overview approaches to modeling the finite rate timescale. As mentioned in Section \ref{sec:PhaseChangeModeling}, the timescale is dependent on both the thermodynamic state $(P, T)$, the composition $(Y_p^c)$, and the interfacial surface area $\Sigma$. As discussed in Section \ref{sec:Intro-FiniteRateModels}, most established finite-rate phase change approaches do not incorporate all of the required dependencies, and instead rely on assumptions or correlations to match experiments. In this work, the Hertz-Knudsen model is used because it is generally derived from kinetic theory and is applicable to flows ranging from flashing conditions to droplet evaporation. The form used in this work is given by,
\begin{equation} \label{eq:HertzKnudsen}
    \dot{m}_{g,HK}^c = \frac{\lambda \Sigma(P_{sat} - P)}{\sqrt{2\pi R/W_{p}^cT}}
\end{equation}
$\lambda$ is an $O(1)$ user tuning parameter which can be tuned to experiments and either defined as a function using correlations \cite{fuster2010influence}, or can defined as a constant \cite{lyras2024modelling}.  

The Hertz-Knudsen phase change model is derived in a manner that achieves the generality required to handle flows ranging from both flashing and droplet evaporation systems. What becomes critical for accurately modeling mass transfer with the Hertz-Knudsen model is determining the interfacial surface area. A modeling framework designed to predict the sub-grid interfacial area will be described in Section \ref{sec:SprayModeling}.

\subsubsection{Thermodynamically Bounding Finite Rate Phase Change}
Unlike the HEM model, an issue with the Hertz-Knudsen, D2-law, and HRM (described in \ref{sec:FiniteRateModelForms}) finite rate phase change models is that they are not bound to obey the chemical potential equilibrium condition described in Section \ref{sec:PhaseChangeModeling}. Critically, this means that these finite rate models could unreasonably predict phase change when none should occur, or overshoot the prediction of the HEM (which is unphysical). 

In order to avoid this issue, we propose a new implementation for finite rate phase change models. Instead of a direct mass transfer term, the finite rate phase change models can be formulated as a time scale, which informs the following expression,
\begin{align}
    &{\dot{m_p^c}} = \frac{1}{\tau}\rho \left( Y{_{eq}}_{p}^c -  Y_p^c\right)
\end{align}
where $Y{_{eq}}_{p}^c$ is defined using the HEM, and $\tau$ is defined using a finite rate phase change model. As the Hertz-Knudsen model is based on more fundamental concepts, it will be used as the finite rate phase change model in this work. To have a thermodynamically consistent bound on the phase change rate, we define $\tau$ as,
\begin{equation}
    \tau = \max{\left(\frac{\rho \left(Y{_{eq}}_{p}^c - Y_p^c\right)}{\dot{m}^c_{p,HK}}, \Delta t\right)}
\end{equation}
where in this expression $\dot{m}^c_{p,HK}$ is defined using Eq. \ref{eq:HertzKnudsen}.
With this implementation, the Hertz-Knudsen finite rate phase change model will always predict a physically achievable thermodynamic state. In order to inform the Hertz-Knudsen phase change model, a representation of the subgrid surface area is required. The following section overviews the subgrid modeling approach used in this work.

\section{Sub-grid Spray Modeling} \label{sec:SprayModeling}

As described in Section \ref{sec:FiniteRatePhaseChange}, a critical aspect of modeling the rate of phase change is estimating the interfacial surface area. In this section, the focus is on extending the physical application of the numerical schemes described in Section \ref{sec:NumericalMethod} to consistently include the $\Sigma$ model for LES flows with features far smaller than the grid scale. 

\subsection{Phase Constrained $\Sigma'$ modeling}

The original application of the $\Sigma$ model to the LES setting involves transporting the equation below \cite{anez2019eulerian},
\begin{equation} \label{eq:OriginalSigmaPrimeTransportEQ}
    \frac{\partial \Sigma'}{\partial t} + \frac{\partial \widetilde{u}_i \Sigma'}{\partial x_i} = \frac{\partial}{\partial x_i}\left(\frac{\nu_t}{Sc_t}\frac{\partial \Sigma'}{\partial x_i}\right) + \dot{\Sigma}'_{int}
\end{equation}
where $\Sigma'$ represents the phase interface surface area density that is not resolvable by the mesh size. The total interfacial surface area is obtained by, $\Sigma = \Sigma_{min} + \Sigma'$ where $\Sigma_{min}$ represents the theoretical amount of surface area that is possible in an LES setting. Although the resolved surface area is mathematically defined as the magnitude of the gradient of the volume fraction, $|\nabla \phi_l|$, this definition is not always appropriate when the $\Sigma$ model is used in an LES. To illustrate this point, consider a region of the mesh that is too coarse to resolve the relevant flow structures. Instead of using an interface regularization model, this region relies on a subgrid spray model to distribute the liquid volume over several grid cells. Downstream of the breakup event, it is possible that the mesh resolution becomes sufficient to resolve theoretical spray features. However, since the upstream liquid has already been represented as a diffuse distribution and transported as an unresolved spray, the fine-scale interface structures cannot be easily reintroduced into the simulation where the exact expression for $\Sigma_{min} = |\nabla \phi_l|$ can be used. To address such situations, where interfaces may be partially resolved or artificially diffused even in high-resolution regions, a minimum surface density model, $\Sigma_{min}$, is commonly employed as,
\begin{equation}
    \Sigma_{min} = \frac{2.4}{\Delta}\sqrt{\phi_l(1-\phi_l)}
\end{equation}
where $\Sigma_{min}$ represents the features of the spray that theoretically could be fully resolved by the grid if no spray model was present \cite{chesnel2011large,desantes2020eulerian,gaballa2023modeling}. 

In Eq. \ref{eq:OriginalSigmaPrimeTransportEQ}, the diffusion term on the right-hand-side represents the diffusion of surface area density due to the sub-grid transport. In the literature this is known as the turbulent diffusion flux and is generally based on a gradient law closure as shown in Eq. \ref{eq:OriginalSigmaPrimeTransportEQ} \cite{lebas2009numerical}.  The final term on the right-hand-side, $\Sigma'_{int}$, consists of source and sink terms to represent physical spray events that include flashing, dense turbulent breakup, dilute breakup, and coalescence. The model forms for the source terms will be described in Section \ref{sec:DenseSpraySourceTerms} and Section \ref{sec:DiluteSpraySourceTerms}.

The transport equation in Eq. \ref{eq:OriginalSigmaPrimeTransportEQ} has been used in past work to successfully model the transport and production of surface area throughout Eulerian simulations \cite{lebas2005coupling, anez2019eulerian}. In general, the $\Sigma'$ equation is a scalar transport equation that should consistently follow the Eulerian flow field produced by the Navier-Stokes equations. In the absence of source terms, the quantity $\Sigma'$ should only physically exist within the volume of the gas phase, since it represents the liquid surface area. A pure liquid core should not contain $\Sigma'$, as there is no surface area in a pure phase. Furthermore, within a computational cell that contains liquid surface area, all surface area should exist as a confined scalar within the volume occupied by the gaseous phase. In other words, even in the subgrid setting, the liquid droplets, ligaments, and other interfacial features that make up $\Sigma$, will exist within the gaseous volume and either be partially or fully surrounded by the gas phase. However, this desired modeling quality will not be satisfied for the modeled transport equation shown in Eq. \ref{eq:OriginalSigmaPrimeTransportEQ}. In Eq. \ref{eq:OriginalSigmaPrimeTransportEQ}, the transport equation $\Sigma'$ occupies the entire volume of each cell, which has been shown in past work to lead to leakage of confined scalars across phase interfaces \cite{mirjalili2022consistent}. Following the requirements presented in \cite{mirjalili2022consistent}, Eq. \ref{eq:OriginalSigmaPrimeTransportEQ} can be rewritten as
\begin{equation}
    \frac{\partial \Sigma'}{\partial t} + \frac{\partial \widetilde{u}_i \Sigma'}{\partial x_i} = \frac{\partial}{\partial x_i}\left[\frac{\nu_t}{Sc_t}\widetilde{\phi_g}\frac{\partial}{\partial x_i}\left(\frac{\Sigma'}{\widetilde{\phi_g}}\right)\right] + \frac{\partial}{\partial x_i}\left(\overline{R_{g,i}^c}\frac{\widetilde Y_p}{\widetilde Y_p^c \overline \rho_p}\frac{\Sigma'}{\widetilde{\phi_g}}\right) + \dot{\Sigma}'_{int}
\end{equation}
where the first term represents the physical diffusion of the transport of $\Sigma'$ by the subgrid flow. In this work, a form of the turbulent diffusion term has been proposed to confine the diffusion of the scalar, $\Sigma'$ to the gas phase. Compared to the diffusion term in Eq. \ref{eq:OriginalSigmaPrimeTransportEQ}, the diffusion term now contains the addition of $\tilde \phi_g$ to limit the diffusion of the scalar quantity, $\Sigma'$, to the volume of the gas. Additionally, a second term is added to the system as a required consistency term to ensure that $\Sigma'$ is properly transported when the turbulent liquid flux spray model is applied to the mass, momentum, and energy equations. Together, these terms allow the transport of $\Sigma'$ to remain fully consistent and leakage-free across phase interfaces. In this work, the gradient diffusion closure of, $\overline {R_{g,i}^c} = \frac{\nu^t}{Sc^t}\nabla\widetilde\phi$, is used to close the unclosed turbulent liquid flux \cite{anez2019eulerian}. Note, unlike the species diffusion subgrid flux ($\overline{J_{p,j}^c}^t$) which acts to model subgrid diffusion between species within phases (intraphase mixing for multi-component mixtures), the turbulent liquid flux model defines the closure of the interaction between phases (interphase mixing between multiphase mixtures).

In the proposed phase-confined transport for the subgrid surface area equation, the final term on the right-hand-size contains sources and sinks for modeling the breakup and coalescence of liquid droplets. Most past works have used a non-linear form for a general source term as \cite{vallet1999modelisation, lebas2005coupling,lyras2019numerical, anez2019eulerian},
\begin{equation}
    S = \frac{\Sigma}{\tau}\left(1 - \frac{\Sigma}{\Sigma_{eq}}\right).
\end{equation}
Although this has been the standard application of the source term in $\Sigma$ modeling and has shown success in multiple applications, for small time-scales (at minimum equal to the simulation time-step) the non-linearity in the definition of $S$ can result in the integration of $\dot{\Sigma}'$ over/under-shooting past the predicted equilibrium value $\Sigma_{eq}$. Instead, a linear relaxation term is used in this work given as,
\begin{equation} \label{eq:SigmaLinearRelaxation}
    S = \frac{1}{\tau}\left(\Sigma'_{eq} - \Sigma'\right).
\end{equation}
The modeling of all spray-physics can be framed using Eq. \ref{eq:SigmaLinearRelaxation} towards a predicted equilibrium value without over/under-shooting $\Sigma'_{eq}$ during time-integration. For different breakup processes, both a time-scale and an equilibrium value are modeled. The following sections will briefly review the source terms explored in this work and the required changes to allow for consistent integration into the current computational framework.

\subsection{Dense Spray Source Terms} \label{sec:DenseSpraySourceTerms}
Within a dense region, multiple sources of surface area production can exist. In this work, two sources of production events will be modeled. The first is a surface area production caused by droplet breakup as a result of turbulence and aerodynamic interactions. The second is a production of surface area resulting from a flashing event within a dense liquid core.

\subsubsection{Turbulent breakup}
For a dense spray with turbulent breakup, the modified approach of \cite{duret2013improving} is used to estimate the equilibrium surface density. The form of this equation is given by,
\begin{equation}
    \Sigma'_{eq,turb_{dense}} = 4\frac{0.5(\overline \rho_l + \overline \rho_g)\widetilde{\phi_l}(1-\widetilde{\phi_l})k_{sgs}}{\sigma We_c}
\end{equation}
where $\sigma$ is the surface tension coefficient and $We_c$ is the critical Weber number which is a user-defined parameter. For the turbulent breakup model, $We_c=1.5$ is used for all cases in this work. In the LES setting, the time-scale used for the relaxation towards equilibrium is the magnitude of the strain-rate tensor as,
\begin{gather}
    \tau_{dense,turb} = \frac{1}{|S_{ij}|}.
\end{gather}
and $k_{sgs}$ is the subgrid turbulent kinetic energy which we define as,
\begin{equation}
    k_{sgs} = \left(\frac{\nu_t}{\Delta C_s}\right)^2
\end{equation}
where $\Delta$ is the local grid size and $C_s = 0.1$ \cite{yoshizawa1985statistically,OpenFOAMv2206_WALE}. This model for dense turbulent breakup has been successfully applied in multiple approaches including RANS and LES \cite{gartner2024novel,gaballa2023modeling}. 

\subsubsection{Flashing}
Although not as commonly explored, explicit models for surface area production in dense zones from flashing have been recently formulated in the RANS context \cite{gartner2024novel}. In the LES context this model can be written as,
\begin{equation}
    \Sigma'_{eq,flashing} = \frac{\overline \rho_L\widetilde\phi \dot{r_f}^2}{2\sigma We_c}
\end{equation}
where $We_c$ is an $O(1)$ user parameter for the flashing model defined using the correction factor described in \cite{gartner2024novel}. Additionally, $\dot{r_f}$ is the growth rate of the nucleation sites at the time of breakup. Although the model contains terms describing bubble growth in a flashing setting, the source term is representative of the surface area production of the liquid/droplet phase. Consistent to what was done in \cite{gartner2024novel}, the timescale can be defined as,
\begin{equation}
    \tau_{flash} = r_f/\dot{r_f}.
\end{equation}
where $r_f$ is the final radius reached before the bubbles merge or breakup. The remaining challenge is predicting the growth rate and the final bubble radius. We follow the same modeling choices as \cite{gartner2024novel}, though we add a thermodynamic consistency check to limit the growth rate to the maximum amount of phase change which is possible by equilibrating the chemical potential. The original model \cite{gartner2024novel} defines the bubble growth rate assuming the heat diffusion limit \cite{prosperetti2017vapor},
\begin{equation} \label{eq:OriginalFlashingSource}
    \dot{r} = \frac{12D^c_g Ja^2}{\pi r}
\end{equation}
where the Jakob number is defined as,
\begin{equation}
    Ja = \frac{\rho_lc_p(T-T_{sat}(P))}{\rho_g L_v}
\end{equation}
with $L_v$ as the latent heat of vaporization and $D_g^c$ is the binary diffusion coefficient for the gas. Similar to \cite{gartner2024novel}, the bubble final radius can be defined using the following expressions,
\begin{equation}
    r_f = r_{crit}r^*, \quad r^* = \left(\phi_d\left[\frac{1}{\eta} - 1 + \frac{\rho_g}{\rho_l}\right]\right)^{-1/3}, \quad r_{crit} = \frac{2 \sigma}{P_{ve} - P}
\end{equation}
where $\phi_d$ is the volume ratio of the dissolved gas in the fluid needed to form nucleation sites, $\eta$ is the maximum bubble growth volume decided as $\eta = 0.74$ which is the largest volume fraction achievable by orderly spherical packing. Additionally, the critical radius $r_{crit}$ is defined using the Young-Laplace equation as the maximum radius that sustains a bubble's internal pressure for a given amount of surface tension. In that term, $P_{ve}$ is the vapor pressure inside the bubble given by the Gibbs-Duhem equation as,
\begin{equation}
    P_{ve} = \exp\left(\frac{P-P_{sat}(T)}{\rho_lRT}\right)P_{sat}(T).
\end{equation}
Although the model described above has been shown to successfully model the flashing of cryogenic nitrogen \cite{gartner2024novel}, similar to the HRM, it is not bound to follow the thermodynamically defined chemical potential equilibrium conditions. To address this issue, instead of using Eq. \ref{eq:OriginalFlashingSource} directly, we can estimate the amount of mass transfer from reaching thermo-chemical equilibrium with bubble growth as,
\begin{align}
    \frac{d}{dt}(\rho_gn_b\frac{4}{3}\pi r^3) &\leq \dot{m} \\
     \rho_g n_b4\pi r^2\dot{r}&\leq \frac{1}{\Delta t}\rho  \left(Y{_{eq}}_p^c - Y_p^c\right) \\ \label{eq:BoundedBubbleGrowth}
     \dot{r} &\leq \frac{\frac{1}{\Delta t}\rho  \left(Y{_{eq}}_p^c - Y_p^c\right)}{\rho_g n_b4\pi r^2}.
\end{align}
According to these modeling assumptions, the maximum growth rate of the bubbles during a flashing process can be bound by the amount of phase change present when reaching chemical equilibrium. Therefore, Eq. \ref{eq:OriginalFlashingSource} can be modified using the results from Eq. \ref{eq:BoundedBubbleGrowth} to consistently predict flashing in zones that have the potential to experience an evaporation or vaporization event as,
\begin{equation} \label{eq:NewFlashingSource}
    \dot{r_f} = \min{\left(\frac{12\mathcal{D} Ja^2}{\pi r_f}, \frac{\frac{1}{\Delta t}\rho \left(Y{_{eq}}_{p}^c - Y_p^c\right)}{\rho_gn_b4\pi r_f^2}\right)}.
\end{equation}

\subsection{Dilute spray source terms} \label{sec:DiluteSpraySourceTerms}

In the dilute spray regime the spray features are assumed to be small droplets which can undergo further secondary breakup, or can coalesce into larger droplets, reducing the overall interfacial surface area. 

\subsubsection{Secondary breakup}

The model for the secondary breakup of droplets is given by \cite{lebas2009numerical},
\begin{equation}
    \Sigma'_{eq,2ndBreak} = \frac{6\widetilde{\phi_g}\overline \rho_gu_{rel}^2}{\sigma We_{c}}
\end{equation}
where $u_{rel}$ is an estimate of the relative velocity, and the equilibrium Weber number is given as $We_c=12.0$ \cite{lebas2009numerical, gartner2024novel} in this work for the secondary breakup model. Furthermore, the breakup timescale is,
\begin{equation}
    \tau_{turb_{dilute}} = 2r/u_{rel} T\sqrt{\rho_l/\rho_g}
\end{equation}
where r is the current estimate of the radius of the droplets calculated with the Sauter mean diameter (SMD), and $T$ is an experimentally determined correlation for secondary droplet breakup \cite{pilch1987use}. The correlation is given by,
\begin{equation}
    T = 1.9(We_{2ndBreak} - 12.0)^{-0.25}(1.0+2.2Oh)^{1.6}.
\end{equation}
where $We_{2ndBreak} = \rho_g u_{rel}^2 2r/\sigma$, and the Ohnesorge number, $Oh = \frac{\mu_l}{\sqrt{\rho_l\sigma 2r}}$. 

\subsubsection{Coalescence}
Lastly, in the dilute region the coalescence of droplets is estimated in the LES setting using particle collision theory and the subgrid turbulent kinetic energy as \cite{lebas2009numerical, gartner2024novel},
\begin{equation}
    \Sigma'_{eq,col} = \Sigma'\frac{6 + We}{6 + We_c}
\end{equation}
where,
$We = \frac{4\phi_l\rho^*k_{sgs}}{\sigma \Sigma'}$,
and 
$\rho^* = 4(\rho_l-\rho)(\phi - 0.5)^2 + \rho$ with a time-scale of,
\begin{equation}
    \tau_{col}= \frac{1}{\Sigma' \sqrt{(2k_{sgs}/3)}}.
\end{equation}
$We_c=12.0$ \cite{lebas2009numerical, gartner2024novel} is used for the coalescence model for all cases in this work.

\subsection{Evaporation source term}
The resulting surface area predictions achieved by incorporating the production and destruction mechanisms for surface area described above directly impacts the phase change rate. Critically, the transfer of mass also impacts the surface area of the liquid phase. In order to keep the surface area predictions consistent, the surface area must be updated when mass is transferred during the phase change process. In general, it is difficult to determine whether mass transfer from phase change increases or decreases surface area. Assuming all phase change due to evaporation and condensation takes place with non-interacting spherical droplets, the following equation is appropriate \cite{lebas2009numerical, gartner2024novel},
\begin{equation} \label{eq:Sevap_droplets}
    S_{evap} = -\frac{2}{3}\Sigma' \dot{m_l^c}/(\rho Y^c_l).
\end{equation}
where evaporation decreases surface area, and condensation increases. Some authors do not include this term as they argue the change of surface area from phase change is implicitly handled by the other source terms \cite{gaballa2023modeling}. In the current study, a modified form of Eq. \ref{eq:Sevap_droplets} is used as,
\begin{equation} \label{eq:Sevap_droplets_mod}
    S_{evap} = -\Sigma' \dot{m_l^c}/\rho.
\end{equation}
which is proportional to the surface area update in Eq. \ref{eq:Sevap_droplets} and captures the general effect of phase change without any robustness issues when $Y_l^c$ gets small. Although we have added Eq. \ref{eq:Sevap_droplets_mod} to be philosophically consistent with phase change impacting surface area, this term is not expected to dominate in relation to the other production/destruction sources of $\Sigma'$ and could be neglected without a strong impact on the results, as discussed in \cite{gaballa2023modeling}.

\subsection{Spray regime indicators}
The integration of the source terms mentioned above are classified into two broad sections,
\begin{equation}
    \dot{\Sigma}'_{int} = S_{Spray} + S_{PhaseChange}
\end{equation}
where the $S_{Spray}$ terms contain the source terms associated with dense turbulent breakup, dilute secondary breakup, and dilute coalescence. However, the $S_{PhaseChange}$ terms contain a production term associated with the flashing process (as this requires mass transfer) and a term associated with evaporation/condensation. 

For both the spray and phase change sections, it is necessary to determine the regime of phase change in order to use the appropriate models.  

\subsubsection{Dense vs. Dilute}

The underlying spray physics for a dense spray are expected to consist of either a connected phase (discrete phase), or a large collection of droplets which would generally correspond to a large liquid volume fraction of the cell. In contrast, the spray physics for the dilute regime can be modeled as a disperse phase that does not occupy a large volume fraction of the cell. In this work, two indicators are used to determine if the spray region consists of a dense or dilute spray.  First, the dense vs. dilute indicator proposed by \cite{gartner2024novel}, which is based on the liquid volume fraction, is used to define the dense region. Instead of solely using the indicator from \cite{gartner2024novel}, an additional indicator based on an a-priori estimate of the Hinze scale can be used to distinguish between dense and dilute regions. The a-priori Hinze scale estimate \cite{khanwale2022breakup} is given by,
\begin{equation}
    \xi = 0.75\left(\frac{\sigma}{\rho_g}\right)^{3/5}\left(\frac{\nu_g^3Re_g^{3/4}/D}{\rho_g}\right)^{-2/5}
\end{equation}
where $D$ is the diameter of the injector for the spray, and $Re_g = \frac{\rho_gU_iD}{\mu_g}$ with $U_i$ as the maximum inlet velocity. So, overall, in this work the dense vs. dilute spray region indicator is given by,
\begin{equation} \label{eq:Dense_vs_Dilute_indicator}
    \Psi = \max\left(\min\left(\max\left(\frac{\phi_l - 0.26}{0.5 - 0.26},0\right),1\right),\Psi_\xi\right)
\end{equation}
where the final term in Eq. \ref{eq:Dense_vs_Dilute_indicator} is given by,
%\begin{equation}
%    \Psi_\xi = \min\left(\max\left(6\phi_l/\Sigma-\xi, 0\right), 1\right).
%\end{equation}
\begin{equation}
\Psi_\xi = \begin{cases}
    1, & \text{if } 6\phi_l/\Sigma > \xi \\
    0,  & \text{otherwise }
\end{cases}
\end{equation}
which compares the current local SMD size (SMD = $6\phi/\Sigma$) to the a-priori estimate of the Hinze scale. In comparison to the original indicator based on the volume fraction, the additional Hinze scale indicator adds a new classification mechanism for the low volume fraction zone of $\phi_l < 0.26$. Even though the volume fraction is low, it is still possible that either large drops or connected structures exist in the subgrid which should be treated with the dense spray models. As such, to use the dilute models we ensure that both the volume fraction and the estimated droplet sizes are small as expected for a droplet which could either be dominated by coalescence or secondary breakup mechanics. All together the spray source terms are written as,
\begin{equation}
    S_{Spray} = \Psi S_{turb_{dense}} + (1-\Psi)(S_{{2ndBreak}} + S_{col}).
\end{equation}

\subsubsection{Flashing vs. Evaporation}
In addition to indicating the dense and dilute regions of the flow, it is important to indicate which regions of the flow undergo phase change processes dominated by flashing, and which regions which are dominated by evaporation. We propose using an indicator based on correlations of flashing vs. mechanical breakup processes proposed by fitting correlations experimentally \cite{cleary2007flashing}. The indicator used in this work can be written as,
\begin{equation}
    \Gamma = \min\left(\max\left(\frac{Ja_c - 55 We_v^{-1/7}}{150 We_v^{-1/7} - 55 We_v^{-1/7}},0 \right), 1\right)
\end{equation}
and the critical Jakob number is defined as \cite{cleary2007flashing},
\begin{equation}
    Ja_c = Ja(1 - \exp\left({-2300\rho_g/\rho_l}\right)).
\end{equation}
Given the indicator for flashing, the addition of the phase change related source terms can be added to the equations as,
\begin{equation}
    S_{PhaseChange} = \Psi\Gamma S_{flashing} + S_{evap}(1-\Psi\Gamma).
\end{equation} 
With the indicators determined, the computational modeling framework is fully defined. The overall system consists of the multiphase multi-component compressible Navier-Stokes equations which are coupled through a finite-rate phase change model to a consistently transported phase-constrained $\Sigma'$ equation for the subgrid interfacial surface area density. Using this governing system, simulations of the classic engine combustion network (ECN) Spray A at non-evaporating and evaporating conditions are used to validate the approach with the ECN experiments.

\section{LES ECN Spray A Simulations} \label{sec:SprayAResults}
To validate the modeling methodologies for both phase change and spray models presented in Sections \ref{sec:PhaseChangeModeling} and \ref{sec:SprayModeling}, the ECN Spray A configuration is studied in this section. The ECN non-evaporating conditions are used to validate the spray modeling terms from Section \ref{sec:SprayModeling}, and the ECN evaporating conditions are used to test both the finite-rate phase change model and the fully coupled system involving both spray characterization and mass transfer. The ECN Spray A case consists of the high-pressure injection of n-dodecane into a quiescent environment of nitrogen \cite{ECN2022f}. The physical parameters for both the evaporating and non-evaporating cases are shown in Table \ref{tab:SprayAPhysicalParameters}. The properties used to determine the state of the mixture in the simulations are in Table \ref{tab:SprayAEOSParameters}. The NASG EOS parameters were defined using the approach described in \cite{boivin2019thermodynamic}.

\begin{table}[hbt!]
    \centering
    \begin{tabular}{l c c}
    \hline
    Thermodynamic Property & Non-evaporating Spray & Evaporating Spray \\
    \hline 
     Injection pressure [MPa] &  150 & 150
    \\[1mm]  
    Injection temperature [K]  & 343 & 363 \\[1mm]
    Ambient temperature [K]  & 303 & 900
    \\[1mm]
    Ambient pressure [MPa]  & 2 & 6
    \\[1mm]
    Ambient composition & N$_{2}$ & N$_{2}$ \\
    \hline
    \end{tabular}
    \caption{Conditions for both non-evaporating and evaporating ECN Spray A cases\cite{ECN2022f}.} \label{tab:SprayAPhysicalParameters}
\end{table}

\begin{table}[hbt!]
    \centering
    \begin{tabular}{l c c c}
    \hline
    Parameter &  Liquid n-dodecane & Vapor n-dodecane & Nitrogen \\[1mm]
    \hline  
    $P_{\infty}$ [Pa]  & $2.8055 \times 10 ^{8}$ & 0.0 & 0.0 \\[1mm]
    $\gamma$ [-] & 1.1595 & 1.021 & 1.4 \\[1mm]
    $C_v$  [$\text{J}$ $ \text{kg}^{-1}\text{K}^{-1}$] & 2147.3 & 2169.5 & 742.87 \\[1mm]
    $\mu$ [Pa-s] & $2.0 \times 10^{-4}$ & $2.0 \times 10^{-5}$ & $1.0 \times 10^{-5}$ \\[1mm]
    b [m$^3$ kg$^{-1}$] & $9.8102 \times 10^{-4}$ & 0.0 & 0.0 \\[1mm]
    q [J kg$^{-1}$] & $-7.4204 \times 10^{5}$ & $-6.61050 \times 10^{5}$ & $-3.0997 \times 10^{5}$ \\[1mm]
    A [-] & \multicolumn{2}{c}{4.10549} & - \\[1mm]
    B [-] & \multicolumn{2}{c}{1625.928} & - \\[1mm]
    C [-] & \multicolumn{2}{c}{-92.839}  & - \\[1mm]
    \hline
    \end{tabular}
    \caption{Simulation parameters for Spray A evaporation and non-evaporation cases.} \label{tab:SprayAEOSParameters}
\end{table}

\begin{figure}[hbt!] 
\begin{center}
\includegraphics[width=\textwidth]{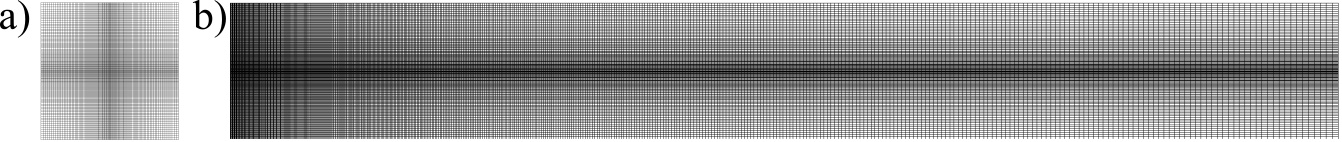}
\caption{Grid 1 used for Spray A non-evaporating case. Panel a) shows the $y$ vs. $z$ plane, and panel b) shows the $x$ vs. $y$ plane. The axial span is from 0 to 20 mm, and the transverse span is from -1.25 to 1.25 mm.  \label{fig:SprayAGrid}}
\end{center}
\end{figure}

As opposed to capturing the full injector geometry, in this work the injection of liquid n-dodecane into the domain is modeled using a uniform inlet matching the injector nozzle diameter of
0.0894 mm \cite{ECN2022a}. Similar to past works, synthetic turbulent forcing is used to trigger the transition to a turbulent jet as opposed to simulating the upstream injector geometry \cite{gaballa2023modeling, desantes2020eulerian}. Additionally, the base inflow conditions use the ECN injector calculator to define time-dependent mass-flow rate obtained from the CMT injection simulator \cite{CMT_InjectionRate}. A grid-convergence study is completed for all quantities of interest for the non-evaporating Spray A case. Stretched Cartesian grids are used for all cases and a sample grid used in this work can be seen in Figure \ref{fig:SprayAGrid}. The grids are defined using the parameters in Table \ref{tab:SprayAGridParameters}. 

\begin{table}[hbt!]
    \centering
    \begin{tabular}{l c c c c}
    \hline
    Name & Axial ER & Transverse ER & Base Spacing [m] & Cells \\
    \hline 
    Grid 1 &  1.005 & 1.025 & $1\times 10^{-5}$ & 6,238,080
    \\[1mm]  
    Grid 2  &  1.005 & 1.025 & $7.1 \times 10^{-6}$ & 10,061,824\\[1mm]
    Grid 3  &  1.005 & 1.025 & $5 \times 10^{-6}$ & 15,564,800
    \\[1mm]
    Grid 4  &  1.005 & 1.025 & $2 \times 10^{-6}$ & 37,945,600
    \\[1mm]
    \hline
    \end{tabular}
    \caption{Grid parameters for setting the grid. The axial and transverse expansion ratios (ER) are constant for all cases. The base spacing used differs with varying resolutions. For Grid 4, a minimum grid spacing of $5\times 10 ^{-6}$ is used for both axial and transverse directions until the spacing using the compounding expansion ratios (starting at the base spacing of $2\times 10 ^{-6}$) compounds more than $5\times 10 ^{-6}$. This allows the smallest grid spacing of $5\times 10 ^{-6}$ to be used for a larger area of cells before increasing to larger sizes. } \label{tab:SprayAGridParameters}
\end{table}

Away from the inlet, all boundaries are specified using NSCBC non-reflective outflow conditions to limit impacts from the boundary on the flow \cite{poinsot1992boundary}. As mentioned in Section \ref{sec:HybridApproach}, the simulations are based on a high-order skew-symmetric KEEP central scheme hybridized with a WENO5Z Godunov scheme \cite{borges2008improved} used across shocks and material interfaces. The details of the Godunov scheme are in \cite{collis2026robust}. A temporal CFL=0.5 is used for all cases.

\subsection{Spray A: non-evaporating case}

\begin{figure}[hbt!] 
\begin{center}
\includegraphics[width=0.8\textwidth]{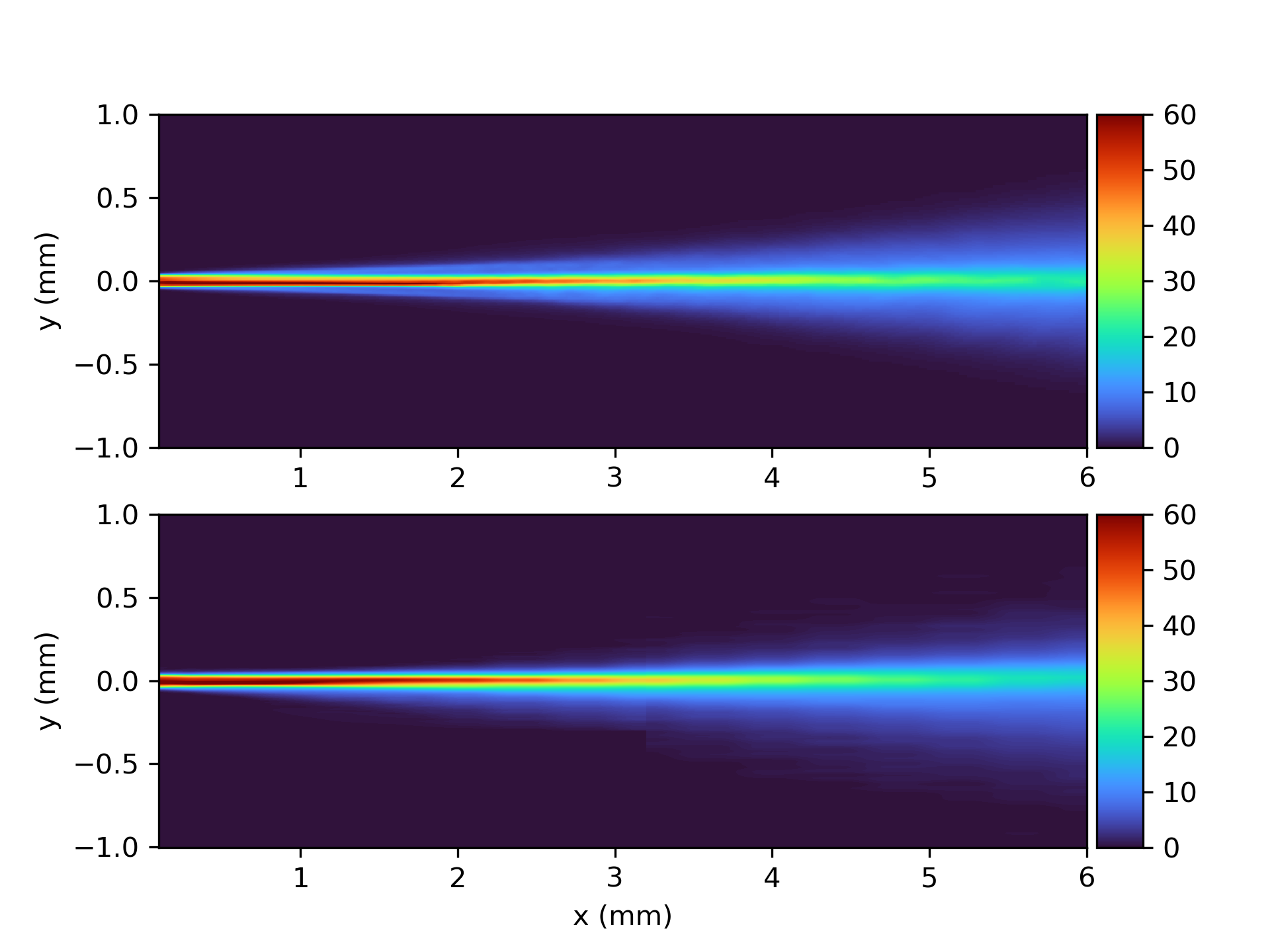}
\caption{Temporal average of the projected mass density [$\mu g/mm^2$] (PMD) in the $x$ vs. $y$ plane. Top row is the LES results achieved with Grid 4 and the bottom row is the experimental result \cite{ECN2022b}. \label{fig:PMD_Average}}
\end{center}
\end{figure}

To validate the application of finite-rate phase change informed by the phase constrained $\Sigma$ model, we compare the predictions of the $\Sigma$ model with those from the ECN Spray A experiment. First, we show the projected mass density (PMD \cite{kastengren2012time}) near-injector results averaged over the flow times ranging from 0.4-0.8 ms. Figure \ref{fig:PMD_Average} shows that the simulations qualitatively capture the spread of the liquid jet in the near-injector region. Visually, at any given axial location ($x$), both the experiment \cite{kastengren2012time,ECN2022b} and the simulation shows a similar magnitude of PMD. Although the experimental results show a less symmetric solution, likely from a bias present in the complex upstream injection of n-dodecane, the radial agreement visually matches between the simulations and experiments.

\begin{figure}[hbt!] 
\begin{center}
\includegraphics[width=0.8\textwidth]{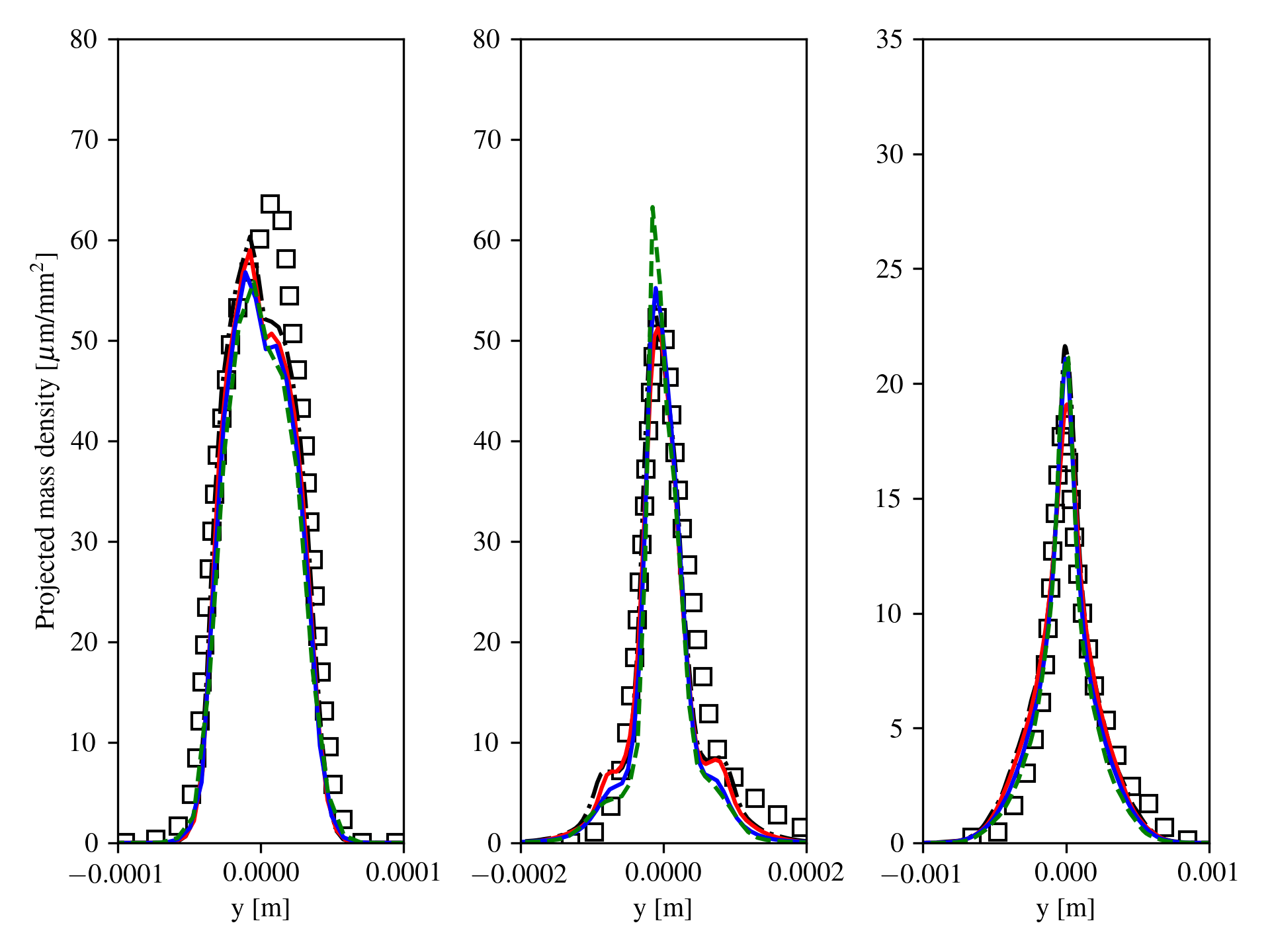}
\caption{Temporally averaged projected mass density [$\mu g/mm^2$] (PMD) along the transverse direction for multiple axial locations. The left panel is for $x = 0.1 $mm, middle panel is for $x = 2$ mm, and right panel is for $x = 6$ mm. Grid 1 (\protect\dashedGreen{}), Grid 2 (\protect\fullBlue{}), Grid 3 (\protect\fullRed{}), Grid 4 (\protect\chainBlack{}), and experiments \cite{ECN2022b}(\protect\markerfour{}). \label{fig:PMD_QuantitativeRadial}}
\end{center}
\end{figure}

A more quantitative analysis for the PMD is shown in Figure \ref{fig:PMD_QuantitativeRadial}, where the experimental profiles are overlaid on the PMD results for varying mesh resolutions. Qualitatively, Figure \ref{fig:PMD_QuantitativeRadial} shows a result similar to Figure \ref{fig:PMD_Average}, where both the magnitude and the overall spread of the PMD are reasonably matched between the experiments and simulations. Additionally, Figure \ref{fig:PMD_QuantitativeRadial} shows how the refinement of the mesh decreases the differences between the spray width in the simulations and experiments. Furthermore, refining the mesh nearly achieves mesh convergence between the two highest resolution runs. The overall agreement between the experimental and simulation PMD provides additional confidence that the underlying LES simulation is representative of the experiment.  

\begin{figure}[hbt!] 
\begin{center}
\includegraphics[width=0.7\textwidth]{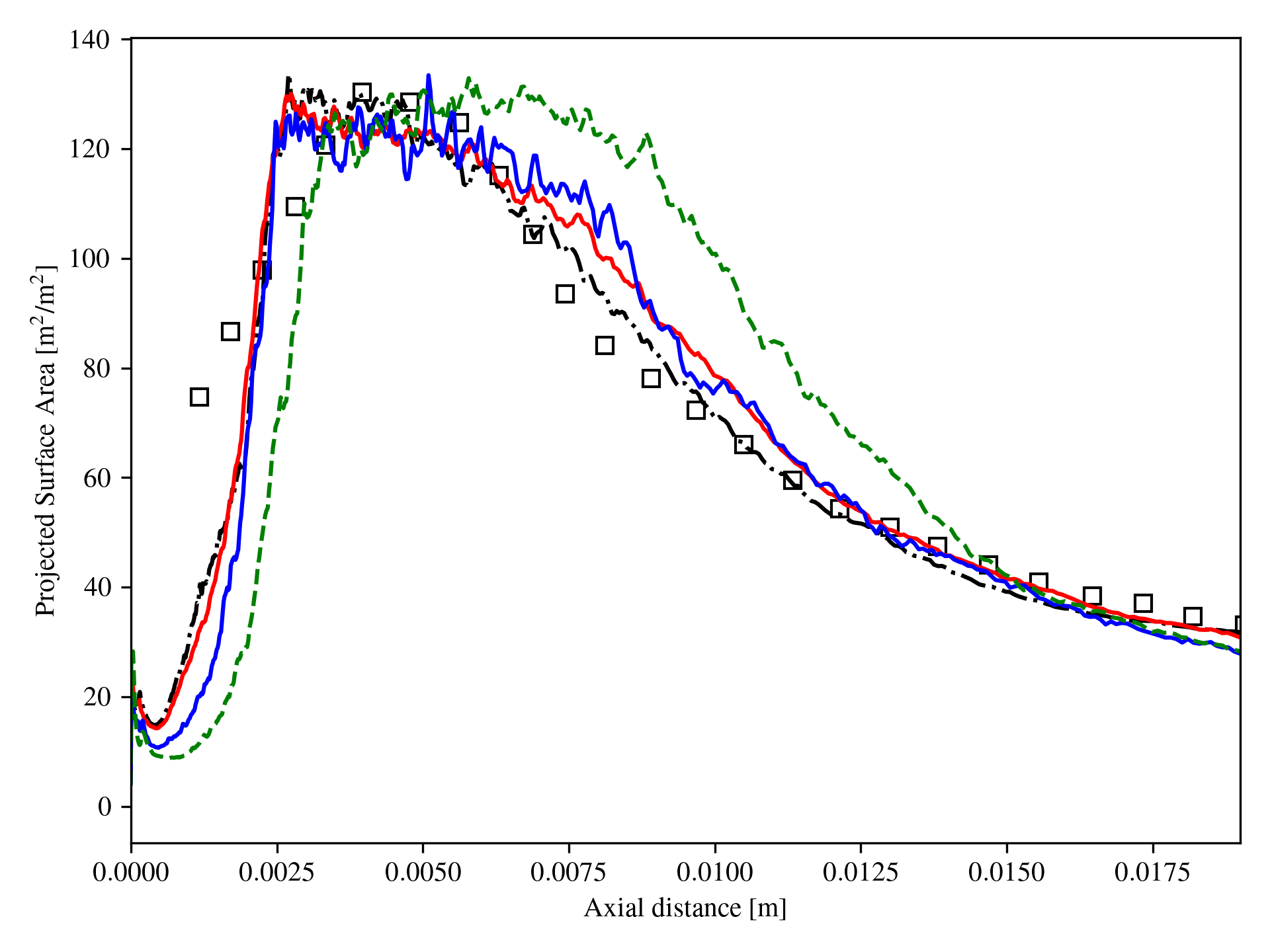}
\caption{Temporal average of the projected surface area [m$^2$/m$^2$] (PSA) along the axial centerline. Grid 1 (\protect\dashedGreen{}), Grid 2 (\protect\fullBlue{}), Grid 3 (\protect\fullRed{}), Grid 4 (\protect\chainBlack{}), and experiments \cite{karathanassis2017comparative}(\protect\markerfour{}). \label{fig:PSA_Axial}}
\end{center}
\end{figure}

\begin{figure}[hbt!] 
\begin{center}
\includegraphics[width=0.8\textwidth]{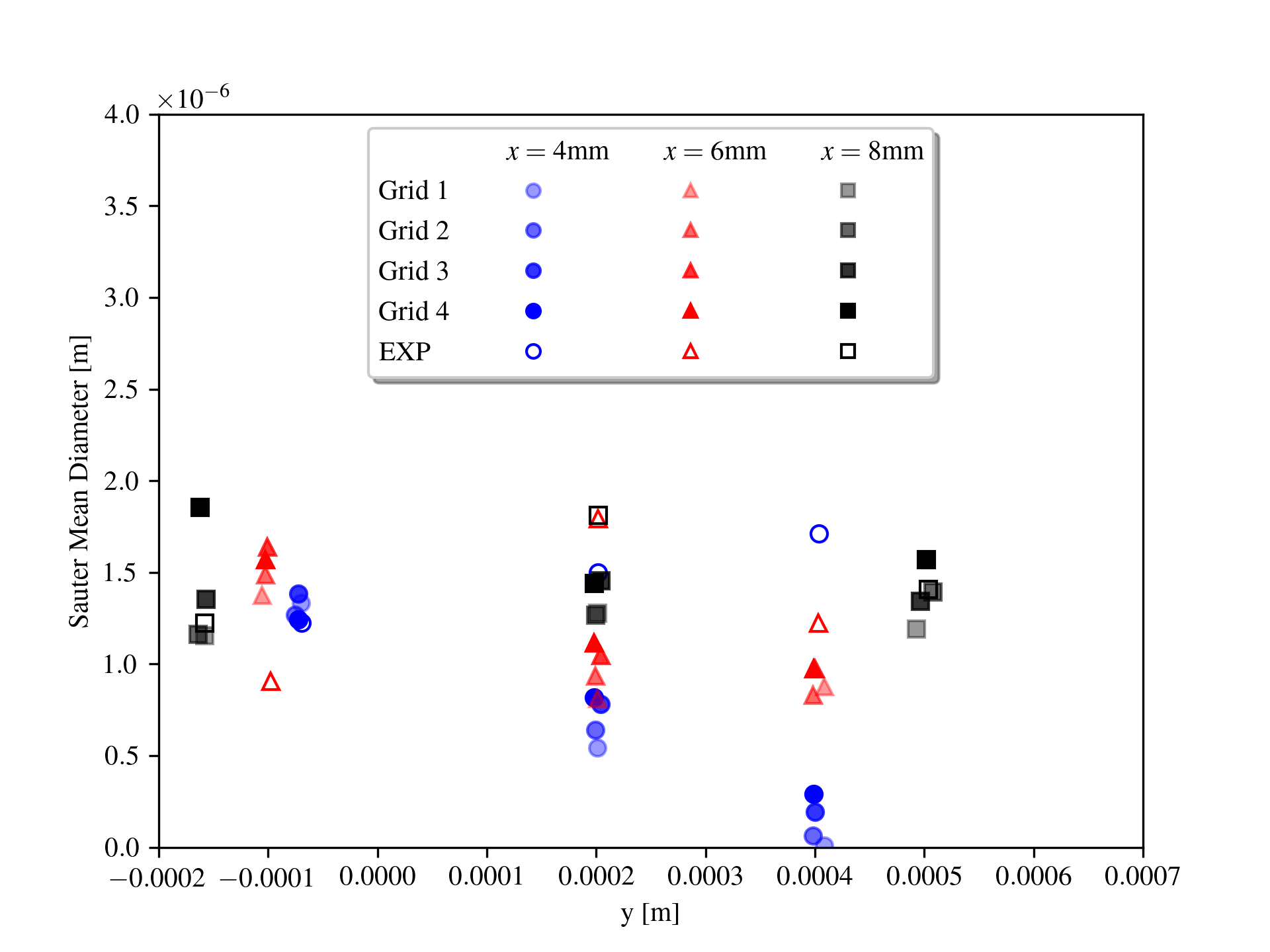}
\caption{Temporal average of the Sauter mean diameter [m] (SMD) at varying transverse and axial locations compared to experiments \cite{karathanassis2017comparative}. \label{fig:SMD_Quantitative}}
\end{center}
\end{figure}

Although the phase change model is active during the simulation, since the flow field does not predict any evaporation (as this is a non-evaporating test-case) the surface area density equation is essentially decoupled from the Navier-Stokes equations. As such, matching the PMD can be viewed as properly simulating the dynamics of the spray with the turbulent liquid flux closure, but it does not validate the surface area predictions of the finite-rate phase change models in the simulations. So, although the results from the PMD analysis are promising, the critical step of validating the spray modeling with the ECN Spray A non-evaporating case requires additional comparison with experimental data. 

To investigate the accuracy of the spray model, Figure \ref{fig:PSA_Axial} shows the projected surface area density predicted by the phase constrained $\Sigma$ model compared to the results from the experiments \cite{karathanassis2017comparative} for varying mesh resolutions. Similar to the results for the PMD, refining the grid in Figure \ref{fig:PSA_Axial} increases the accuracy of the surface area prediction relative to the ECN experiments \cite{karathanassis2017comparative}. Additionally, the two highest grid resolutions are nearly mesh converged. In all cases, the initial rise of the surface area density near the injection of liquid into the domain starts from a lower value than what is observed in the ECN experimental results. The initial rise is dominated by the effects of both spatial resolution and inflow conditions. The qualitative mismatch between simulations and experiments implies that modeling the injection of n-dodecane without the injector geometry does not fully capture the surface area injected into the domain. Even so, further into the domain we see that the spray model properly predicts the overall maximum magnitude of the projected surface area, as well as the reduction in surface area in the far-field region. The agreement between the simulations and experiments provides initial validation on the applicability of the four-equation model coupled with the phase-constrained $\Sigma'$ transport equation to capture the relevant spray physics in the ECN Spray A case.

Further comparisons between the simulation and the experiment can be made using the averaged SMD sizes at varying locations in the domain. Figure \ref{fig:SMD_Quantitative} shows the comparison of the simulation predictions with all grids, as well as the averaged SMD from experiments \cite{karathanassis2017comparative}. In all locations (except the near-field 4mm axial location where the simulations do not predict a wide enough spray), the predicted average SMD sizes from the simulations are in the same range as the experiments. In particular, some near-field locations, including the 4 mm axial location at -0.1 mm radial location, and some far field locations including the 8 mm axial at 0.5 mm radially, obtained close agreement between simulations and experiments. 

\subsection{Spray A: evaporating case}

Following the validation of the non-evaporating Spray A case, the fully coupled system is tested with the evaporating case. As near grid-converged solutions are obtained with the finest grid studied in the non-evaporating case, the fine grid resolution is used for the evaporating ECN Spray A case. The additional physics present in this case includes the finite-rate phase change model coupled with the predictions from the $\Sigma$ model, as well as the potential for flashing in the diesel jet changing the surface area predictions.  

\begin{figure}[hbt!] 
\begin{center}
\includegraphics[width=\textwidth]{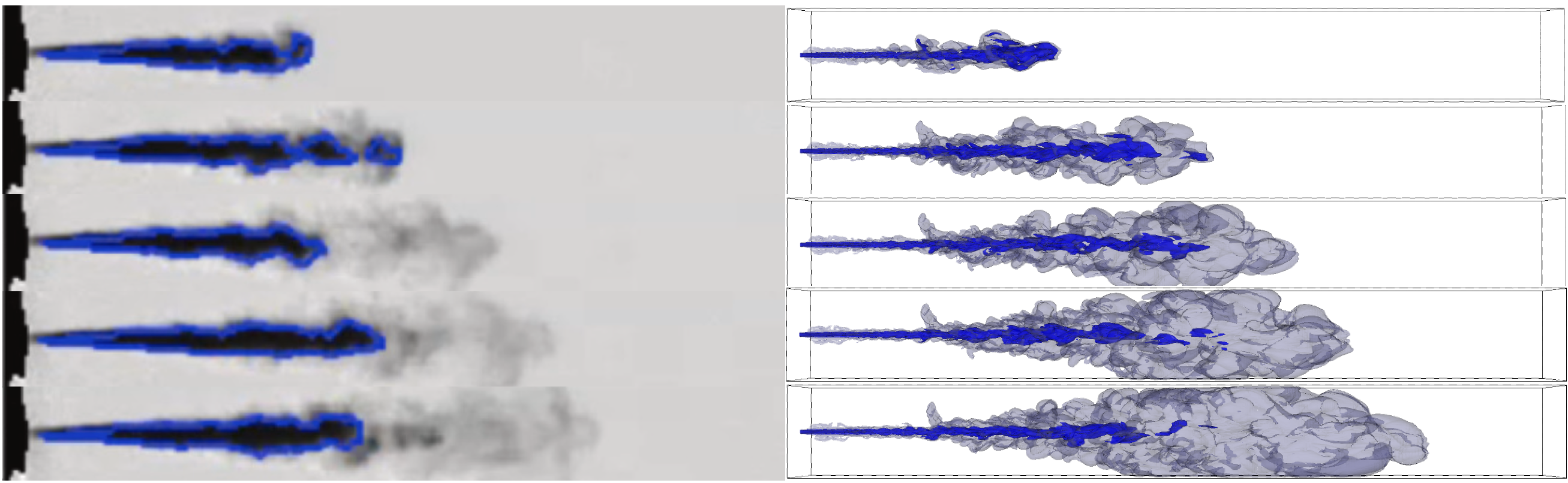}
\caption{Qualitative comparison between evaporating Spray A experiments \cite{manin2012sp2}(left) and LES simulations (right) over time. From top to bottom, the times shown in both experiment and simulation are, $30 \mu s$, $50 \mu s$, $70 \mu s$, $89.9 \mu s$, $109.9 \mu s$. Blue contour represents liquid penetration using volume fraction of $0.15\%$ and gray represents gaseous n-dodecane penetration with mass fraction of $0.1 \%$. \label{fig:SprayA_LiquidGasPenetration}}
\end{center}
\end{figure}

\begin{figure}[hbt!] 
\begin{center}
\includegraphics[width=0.6\textwidth]{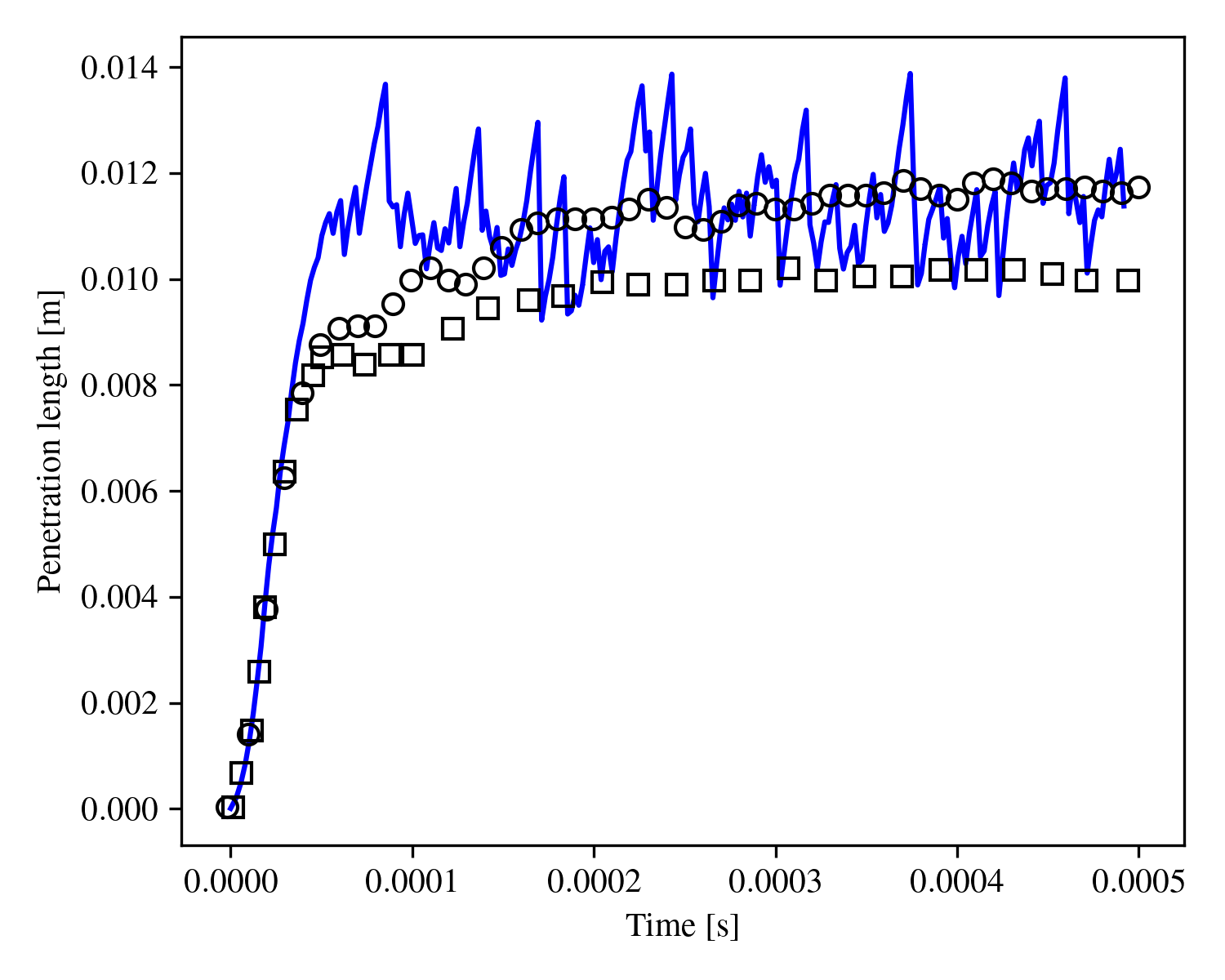}
\caption{Quantitative comparison of liquid penetration into the domain compared to two experimental measurement techniques. LES (\protect\fullBlue{}), experimental DBI \cite{manin2012sp2,ECN2022d} (\protect\markerone{}), and experimental Mie-scattering \cite{bardi2012engine, ECN2022e} (\protect\markerfour{}). \label{fig:SprayA_LiquidPenetration}}
\end{center}
\end{figure}

A qualitative comparison with experiments \cite{manin2012sp2, bardi2012engine, ECN2022d, ECN2022e} can be seen in Figure \ref{fig:SprayA_LiquidGasPenetration}, where the liquid n-dodecane penetration length is shown with a blue iso-contour and the full extent of the gaseous n-dodecane is shown in the gray. The qualitative agreement between both the liquid penetration length and the gas extent into the domain is visually close to the experimental results. In the simulation, the liquid contour is defined using the 0.15\% liquid volume fraction to be consistent with the experimental measurements following Mie-Scattering theory \cite{pickett2015uncertainty}, and the gas n-dodecane contour is defined using the 0.1\% gas mass fraction following experimental guidelines \cite{ECN2022c}. The overall visual behavior of both the liquid core throughout time, as well as the penetration of the gas cloud, match reasonably well between the experiments and fully coupled simulations. 

A more quantitative comparison between experiments and simulation with the liquid penetration length can be seen in Figure \ref{fig:SprayA_LiquidPenetration}. Here, two experimental techniques are shown to classify the extent of the liquid spray into the domain. In addition, the two experimental techniques included in Figure \ref{fig:SprayA_LiquidPenetration} provide a sense of the uncertainty range associated with the experimental readings. After an initial over-penetration of the liquid in the simulation, the majority of the simulation results fall within the experimental measurements. Overall, the agreement in liquid penetration length over time between the experiments and simulations validates the approach of using the four-equation model described in \cite{collis2026robust} coupled with the proposed thermodynamically bounded finite-rate phase change model based on the surface area predictions to capture both complex finite-rate phase change and spray dynamics. 

\begin{figure}[hbt!] 
\begin{center}
\includegraphics[width=0.6\textwidth]{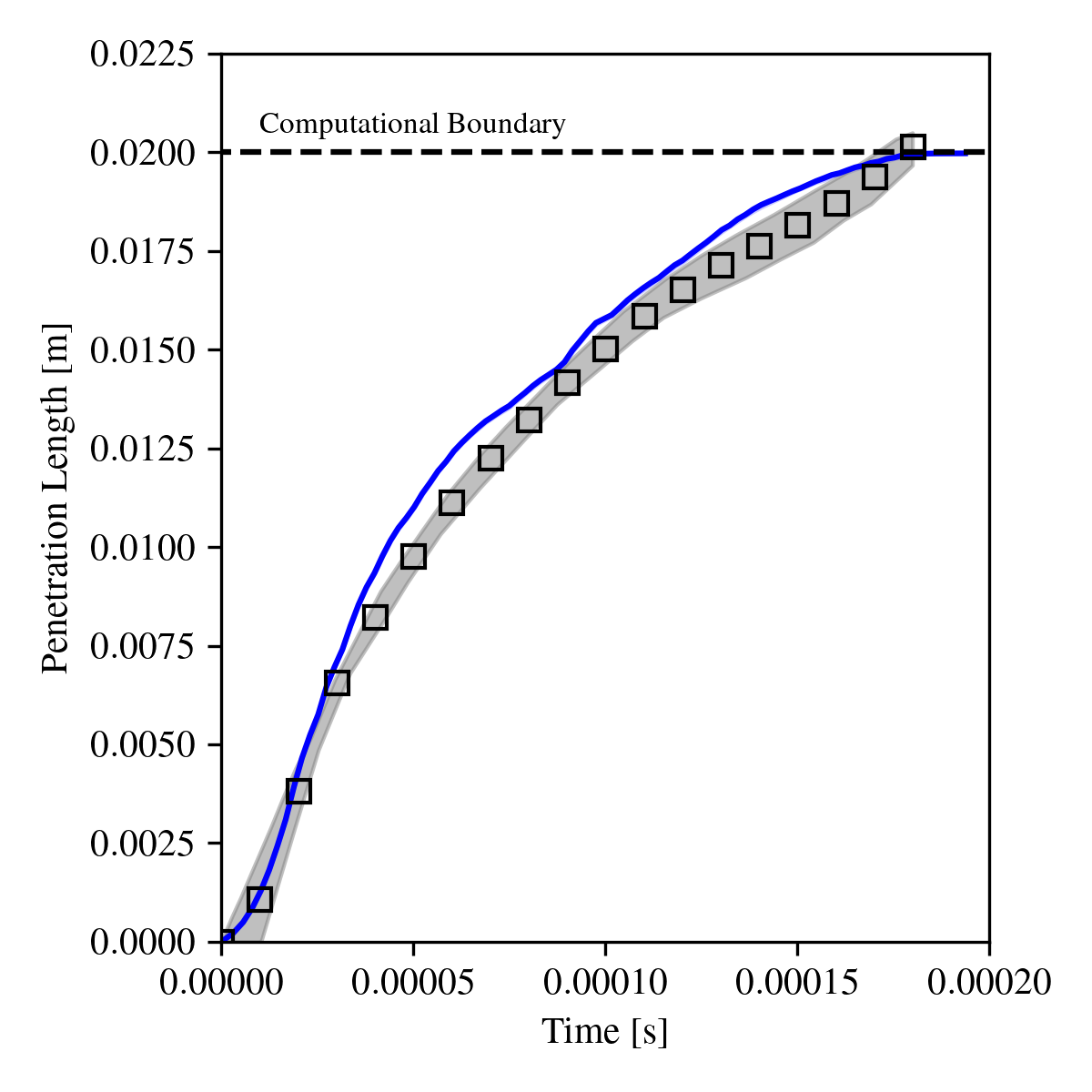}
\caption{Comparison of vapor n-dodecane penetration over time between experiments with confidence intervals, and LES simulations. LES (\protect\fullBlue{}), experimental results \cite{bardi2012engine,ECN2022e} with confidence bounds (\protect\markerfour{}).\label{fig:SprayA_GasPenetration}}
\end{center}
\end{figure}

The final quantity of interest to compare with experiments is the gas penetration length. The n-dodecane gas phase solely exists due to the finite-rate phase change of the injected liquid. The extent of the gas phase n-dodecane into the domain over time can be seen in Figure \ref{fig:SprayA_GasPenetration}. In the simulation, there is a small over-estimation of the gas-penetration throughout time compared to the experimental measurements. As indicated by \cite{gaballa2023modeling}, the over-prediction of gas transport could be attributed to using the ideal gas law to represent a n-dodecane at high temperature and pressure, when in reality it behaves as a supercritical fluid. Future investigation and improvement of the simulation framework would involve increasing the fidelity of the equation of state to allow for more accurate representations of the supercritical n-dodecane. 

Overall, using a thermodynamically bounded and consistent finite-rate phase change model coupled with the predictions of subgrid surface area from the phase-constrained $\Sigma'$ equation provided reasonable agreement with experiments for both the non-evaporating and evaporating ECN spray A cases. Future investigations can include increasing the fidelity of the equation of state, as well as using the hybridization of the interface regularization terms for resolved regions with the surface area representation of the subgrid zones.

\section{Conclusion} \label{sec:Conclusion}

This work presented an extension of the thermodynamically consistent four-equation model presented in \cite{collis2026robust} to include finite-rate phase change and subgrid spray models. Critically, the thermal and mechanical equilibrium assumptions of the four-equation model enable a simple and cost-effective modeling framework for modeling spray dynamics compared to the fully non-equilibrium multiphase model. To achieve agreement with experimental results, a new finite-rate phase change model was formulated to remain thermodynamically bounded by the equilibration of the chemical potential to remove the ability for the phase change model to under/over-predict unphysical mass transfer. Additionally, the finite-rate phase change model was informed using a newly proposed phase-constrained form of the Eulerian $\Sigma$ spray model. The $\Sigma$ transport equation included models to capture the breakup and coalescence processes expected in sprays containing dense and dilute zones and a newly proposed thermodynamically bounded model for flashing in the LES setting. The full simulation framework was validated against the evaporating and non-evaporating ECN Spray A experiments by showing excellent agreement with multiple experimental diagnostic measurements. Future work can include adding mass transfer from chemical reactions to study spray combustion, extending the equation of state to higher fidelity models (e.g. tabular EOS), and investigating hybridizing the spray in the LES setting with regularization of the multiphase interface in resolved zones.  

\appendix

\section{Hybrid Numerical Approach} \label{sec:HybridApproach}

\begin{figure}[hbt!]
\begin{center}
\includegraphics[width=0.6\textwidth]{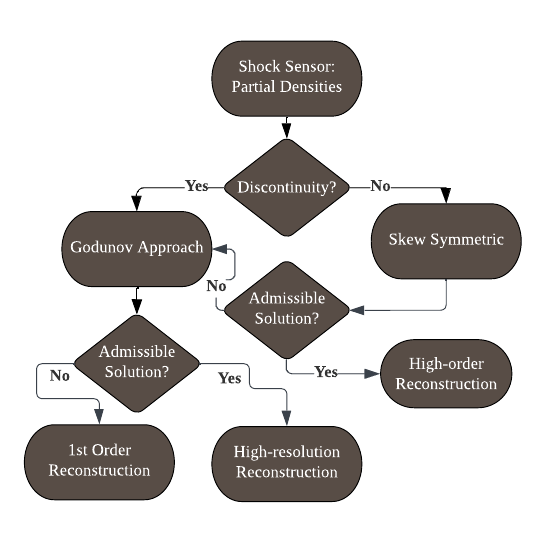}
\caption{Flowchart of hybrid discretization scheme for the convective terms with positivity preserving limiter. \label{fig:ConvectiveFlowchartHybridization}}
\end{center}
\end{figure}

The complete algorithm for the hybrid scheme used in this work can be seen in Figure \ref{fig:ConvectiveFlowchartHybridization} and is described in additional detail in \cite{collis2025robust}. First, the computational cells are labeled using the indicator from Section \ref{sec:HybridDensityIndicator}. After finding a convective flux with either the Godunov approach described in \cite{collis2026robust} or with the skew-symmetric approach described in  \cite{kennedy2008reduced,Kuya2021,Jain2022KEEP}, the solution is checked to be admissible \cite{wong2022positivity}. If the skew-symmetric scheme returns a non-admissible solution, the Godunov approach is attempted before going directly to a 1st-order HLLC reconstruction \cite{toro2013riemann} to ensure that the highest fidelity solution is obtained throughout the domain.

\subsection{Density Shock Sensor} \label{sec:HybridDensityIndicator}

Before applying either the Godunov approach or the skew-symmetric scheme to a face, a density-based indicator is evaluated over the domain to label which cells require the more dissipative Godunov approach. The indicator is based on the evaluation of the sub-stencil smoothness parameter of the TENO6-A scheme \cite{fu2018new}. As described in \cite{fu2018new}, the cutoff value $C_T$ is changed locally in space and time for the adaptive TENO6 scheme. In this work, the density indicator is adapted locally using,

\begin{equation}
    C_T = 10^{-[12.5-8.5(1-d)]}
\end{equation}
where $d = (1-\bm\Phi)^{10}(1+10\bm\Phi)$ and $\bm \Phi$ is the maximum value of the modified Ducros sensor over the stencil with points $i=0,...i_s$, given by,
\begin{equation}
\begin{split}
    &\bm \Phi = \max_{i=0}^{i_s}{\Phi^{(i)}} \\
    &\Phi^{(i)} = \frac{|\vec{\nabla} \cdot\vec{u}|^{(i)}}{\sqrt{\left((\vec{\nabla} \cdot\vec{u})^2\right)^{(i)} + \left(|\vec{\nabla} \times\vec{u}|^2\right)^{(i)} + \epsilon}}. 
\end{split}
\end{equation}
For a given direction and species equation, the TENO6-A scheme is evaluated as smooth using,
\begin{equation} \label{eq:hybrid_cuttoff_expr}
    \frac{\lambda_i}{\sum_{j=1}^3\lambda_j} < C_T.
\end{equation}
where $\lambda_i$ is the same smoothness indicator described in \cite{fu2018new}. The final cell label is determined based on the union of Eq \ref{eq:hybrid_cuttoff_expr} across all species equations. If any sub-stencil for any species equation is not smooth, the Godunov approach is used for both the current cell and its immediate neighbors in that direction (not including diagonal).

\section{Finite-rate Phase Change Model Forms} \label{sec:FiniteRateModelForms}
\subsubsection{Homogeneous Relaxation Model}

As described in Section \ref{sec:Intro-FiniteRateModels}, unlike the instantaneous thermo-chemical equilibrium model (HEM), the Homogeneous Relaxation Model (HRM) defines finite evaporation rate using a model form with parameters informed from experiments. The original HRM was proposed to approximate flashing experiments \cite{downar1996non}. It is given by the form,
\begin{equation} \label{eq:HRM}
    \dot{m}_g^c = \rho Y_g^c\frac{h_l^c - h^c{_l}_{sat}}{h^c{_g}_{sat}-h^c{_l}_{sat}}\frac{1}{\Theta}
\end{equation}
with $h^c_l$ as the liquid enthalpy, $h^c_g$ as the gas enthalpy, and a time-scale,
\begin{equation} 
    \Theta = \Theta_0\epsilon^\gamma\psi^\theta
\end{equation}
where $\psi$ was originally proposed for low-pressure experiments as,
\begin{equation}
    \psi = \frac{P_{sat} - P}{P_{sat}}
\end{equation}
and high-pressure experiments as,
\begin{equation}
    \psi = \frac{P_{sat} - P}{P_c - P}
\end{equation}
with $P_c$ as the pressure at the critical point. Furthermore, in the HRM $\epsilon$ is the void fraction given by $1-\phi$ where $\phi$ is the volume fraction of the liquid. All other parameters in Eq. \ref{eq:HRM} are constants that have been determined by fitting the evaporation rates to experiments. For example, for low-pressure flashing flows \cite{downar1996non} proposed, $\Theta_0 = 6.51 \times 10 ^{-4} s$, $\gamma = -0.257$, and $\theta = -2.24$, and for high-pressure flashing flows \cite{downar1996non} proposed, $\Theta_0 = 3.84 \times 10^{-7}$, $\gamma = -0.54$, and $\theta = -1.76$.

\subsubsection{Droplet Evaporation (D2 Law)}

For droplet dominated flows, correlations with experimental measurements can be defined to capture the D2 evaporation law. As discussed in Section \ref{sec:Intro-FiniteRateModels}, a popular correlation from \cite{abramzon1989droplet} uses thin film theory to approximately capture the effects of heat transfer through a droplet throughout the evaporation process. The overall expression for the droplet mass transfer rate is,
\begin{equation}
    \dot{m}_g^c = 2\pi\rho_gD_gr_sSh^*\ln(1+B_M)
\end{equation}
where $\rho_g$ is the density of the gas mixture in the film, $D_g$ is the binary diffusion coefficient of the gas, $r_s$ is the droplet radius, and $Sh^*$ is a modified Sherwood number given by,
\begin{equation}
    Sh^* = 2 + (Sh_0 - 2)/F_M
\end{equation}
with $Sh_0 = 2 + 0.552 Re^{1/2} Sc^{1/3}$, where $Re$ is the droplet Reynolds number and $Sc$ is the Schmidt number. Additionally, $F_M$ is defined as,
\begin{equation}
    F_M = (1 + B_M)^{0.7}\frac{\ln(1+B_M)}{B_M}
\end{equation}
where $B_M = \frac{Y_l^c{_s} - Y_l^c{_\infty}}{1 - Y_l^c{_s}}$ and $Y_l^c{_s}$ is the mass fraction of the liquid on the surface of the droplet, and $Y_l^c{_\infty}$ is the mass fraction of the liquid far from the surface. A critical addition of this model compared to the HRM formulation is the dependence on the radius of the droplet. Although this increases the accuracy of this model for droplet evaporation, it requires a representation of the droplet sizes throughout a simulation which is not present in many solvers that use Eulerian representations for the liquid. Furthermore, this model is only applicable for droplets that are undergoing evaporation and does not generalize to all flows, including boiling, flashing, or condensation dominated flows.

\bibliographystyle{elsarticle-num}
\bibliography{references}

\end{document}